\documentclass[12pt]{iopart}
\usepackage{iopams}  
\usepackage{color}
\usepackage{graphicx}

\begin{document}

\title[Anisotropy-driven topological quantum phase transition in magnetic impurities]
{Anisotropy-driven topological quantum phase transition in magnetic impurities}

\author{G. G. Blesio}
\address{Instituto de Física de Rosario (CONICET) and Facultad de Ciencias Exactas, Ingeniería y Agrimensura, 
Universidad Nacional de Rosario, Rosario, Argentina}

\author{L. O. Manuel}
\address{Instituto de Física de Rosario (CONICET) and  Facultad de Ciencias Exactas, Ingeniería y Agrimensura, 
Universidad Nacional de Rosario, Rosario, Argentina}

\author{A. A. Aligia}
\address{Instituto de Nanociencia y Nanotecnolog\'{\i}a CNEA-CONICET, GAIDI,
Centro At\'{o}mico Bariloche and Instituto Balseiro, 8400 Bariloche, Argentina}

\ead{manuel@ifir-conicet.gov.ar}
\vspace{10pt}

\begin{indented}
\item[]December 2023
\end{indented}

\begin{abstract}
A few years ago, a topological quantum phase transition (TQPT) has been found in Anderson and Kondo 2-channel 
spin-1 impurity models that include a hard-axis anisotropy term $DS_z^2$ with $D > 0$. The most remarkable 
manifestation of the TQPT is a jump in the spectral density of localized electrons, at the Fermi level, from 
very high to very low values as $D$ is increased. If the two conduction channels are equivalent, the transition 
takes place at the critical anisotropy $D_c \sim 2.5\; T_K$, where $T_K$ is the Kondo temperature for $D=0$. 
This jump might be important to develop a molecular transistor. 
The jump is due to a corresponding one in the Luttinger integral, which has a topological non-trivial value 
$\pi/2$ for $D > D_c$. 
Here, we review the main results for the spectral density and highlight the significance of the theory for the 
interpretation of measurements conducted on magnetic atoms or molecules on metallic surfaces. 
In these experiments, where $D$ is held constant, the energy scale $T_K$ is manipulated by some parameters. 
The resulting variation gives rise to a differential conductance $dI/dV$, measured by scanning-tunneling 
spectroscopy, which is consistent with a TQPT at an intermediate value of $T_K$.
For non-equivalent channels and non-zero magnetic field, the topological phase is lost but still a peculiar
behaviour in the spectral density is obtained which agrees with experimental observations.
We also show that the theory can be extended to integer spin $S>1$ and two-impurity systems. 
This is also probably true for half-integer spin and non-equivalent channels in some cases.
\end{abstract}
\section{Introduction}

\label{intro}

Systems with individual magnetic atoms~\cite{yang19} or molecules~\cite{aradhya13,cuevas10,mathew18,kugel18,evers20} 
on metallic surfaces are being studied extensively in the last years due to their peculiar properties and potential 
application in spintronics and molecular electronics. A fundamental component within an integrated circuit is the 
transistor, which can be realized in these magnetic impurity systems through the controlled switching of the electric 
current by varying some physical parameters~\cite{mathew18}. 

A realistic analysis of magnetic impurity systems requires taking into account their potential multiorbital 
nature and hybridization with more than one conduction channel. Indeed, a plethora of electronic states, 
ranging from Landau-Fermi liquid to singular Fermi liquid and non-Fermi liquid, emerges depending on the 
specific value of the spin of the magnetic impurity and the number of conduction channels~\cite{nozieres80}. 
An important, although often overlooked, physical ingredient of multiorbital systems is the single-ion magnetic 
anisotropy $D S_z^2$ due to the spin-orbit coupling, enhanced in the low-symmetry arrangements of atoms or molecules 
on surfaces. 

As an example of the relevance of single-ion anisotropy, some years ago, it has been found that the conductance of a 
system composed of a Ni impurity in a Au chain doped with oxygen has a jump as a function of the anisotropy $D$ of the 
spin 1 of the Ni atom, suggesting that the system could act as a transistor~\cite{blesio18,blesio19}.
The underlying model is the 2-channel spin-1 Anderson model with anisotropy (2CS1AMA) or its integer valent limit,
the anisotropic two-channel spin-1  Kondo model (A2CS1KM). 
Both models are described in Section \ref{models}. With such models, it can be seen that the jump is due to a 
local topological quantum phase transition (TQPT) between two phases that differ in the value of the so-called 
Luttinger integral $I_L$, whose zero value had been for decades a hallmark of a Fermi liquid~\cite{luttinger60b}. 
For low $D/T_K$ where $T_K$ is the Kondo temperature for $D=0$, the system is in the topologically trivial phase with 
$I_L=0$, characterized by a large spectral density of localized electrons at the Fermi level and large conductance at 
low temperatures and bias voltage. For large $D/T_K$ the system is topologically non-trivial with $I_L=\pi/2$, with a 
pronounced dip in the conductance and spectral density of localized electrons at low energies. This phase has been 
called \textquotedblleft non-Landau \textquotedblright Fermi liquid, because it cannot be adiabatically connected to a 
non-interacting system for which $I_L=0$~\cite{blesio18,luttinger60b}. 

For degenerated channels and zero magnetic field $B=0$, the transition takes place at the critical anisotropy 
$D_c \sim 2.5\;T_K$.  However, the TQPT still exists for $B \neq 0$ or inequivalent channels with $B = 0$~\cite{zitko21}. 
For non-equivalent channels and $B \neq 0$, the system is a topologically trivial conventional Fermi liquid, but a crossover 
from a dip to a peak can be induced by modifying a parameter like $B$~\cite{zitko21} and has been actually observed, as
explained in Section \ref{fepc}.

Previous studies have identified a variety of local QPTs in impurity systems. 
Among the most notable are the transition from the conventional Kondo state to the non-Fermi liquid state in 
two-channel Kondo systems and the Kondo to local moment transition in the 
pseudogap Kondo model (see, for example, Refs.\cite{pusti04, fritz04, vojta06} and references therein). 
However, the nature of our transition is different. 
In our case, the transition takes place between two topologically different Fermi liquids~\cite{blesio18,zitko21}, characterized by different Luttinger integrals $I_L$ --which can be reexpressed as a topological winding number~\cite{blesio18}-- implying necessarily that one of these liquids 
cannot be adiabatically connected to a non-interacting system. 
This liquid has been called a topological or a ``non-Landau'' one to distinguish it from a conventional Fermi liquid. 
Similar QPTs to ours have also been observed in two-impurity systems, where each impurity is coupled to its own conduction 
band, and the two impurities interact through both an antiferromagnetic exchange interaction and a direct Coulomb interaction~\cite{curtin18,nishikawa18}.

In two-channel spin-1/2 models there is a QPT from a Fermi liquid to a non-Fermi liquid~\cite{pusti04}, whose  
Luttinger integral can take a continuous set of values, depending on the occupation of the impurity~\cite{mitchell11}. 
In the pseudogap case~\cite{fritz04}, the transition takes place between a conventional Fermi liquid and a 
local moment phase, consisting in a  Fermi liquid with an uncoupled spin 1/2. 
In the one-channel singlet-triplet models (which hybridizes a configuration with a singlet and a triplet with another one with a doublet)~\cite{allub95,logan09,roura10,florens11}, the QPT separates a conventional Fermi liquid and a singular Fermi liquid, characterized by a non-analytical low frequency self-energy, proper of an underscreened Kondo effect. 
The singular Fermi liquid is characterized by a discrete non-zero $I_L$~\cite{mitchell11,logan09,mitchell13,logan14}. However, as for non-Fermi liquids, the Friedel-Langreth sum rules discussed in Section \ref{res} are not expected to apply.
A spin $S=3/2$ with anisotropy $D$ interacting with two conduction channels undergoes a QPT at $D=0$ where a system is a singular Fermi liquid. This critical point separates a Fermi liquid phase with a decoupled spin for $D<0$ and a non-Fermi liquid phase for $D>0$ \cite{dinapoli13,dinapoli14}.

In this article, we review the main results concerning the above mentioned local TQPT. 
While the original model was proposed 
for a particular system that has not been realized up to date~\cite{dinapoli15}, 
we show that several systems with magnetic atoms or molecules 
on metallic surfaces can be actually described by the 2CS1AMA, or the simpler A2CS1KM, providing a consistent and unified 
description of several experiments. Alternatives theories, sometimes inconsistent between them or physically unjustified, 
have been proposed to interpret the outcome of those experiments. 
This is probably due to the fact that the theory is rather recent, and that to describe the TQPT, a 
very accurate technique like the numerical renormalization group (NRG) is required, which is computationally expensive 
for two or more channels. So far, no other technique has been able to capture the TQPT. 
We also show that the theory can be extended to a larger spin and two-impurity systems.

The paper is organized as follows. In Sec.~\ref{models} we introduce the two models, Kondo and Anderson, for the spin-1 
impurity coupled with two conduction channels. In Sec.~\ref{res} the NRG predictions for the localized electron spectral density 
are presented and discussed in the context of the TQPT. 
In Sec.~\ref{equiv} we show that the models we studied  can be mapped onto those of two $S=1/2$ impurities coupled through an (in general anisotropic) exchange interaction between them. 
In Sec.~\ref{exte} the theory is generalized 
for $S > 1$.
In Sec.~\ref{comparison}, the differential conductance ($dI/dV$) measurements for five different systems are discussed and compared with the theoretical 
calculations for $S=1$: \ref{fepc}) FePc on Au(111), ~\ref{mnpc}) MnPc on Au(111), ~\ref{nc}) nickelocene on Cu(100), ~\ref{trish}) 
Fe atoms on MoS$_2$/Au(111) and \ref{porp}) Fe porphyrin molecules on Au(111).  Finally, we conclude with a brief summary in Sec.~\ref{summary}.

\section{Models}

\label{models}

The simplest model to describe the local TQPT is the anisotropic two-channel spin-1  Kondo model (A2CS1KM). 
It takes the form
\begin{equation}
H_{K} = \sum_{k\tau\sigma}\varepsilon_{k\tau}c_{k\tau\sigma}^{\dagger}c_{k\tau \sigma} + 
\sum_{k\tau \sigma \sigma'}\frac{J_{K}^{\tau}}{2}c_{k\tau \sigma}^{\dagger }\vec{\sigma}_{\sigma\sigma'}
c_{k\tau\sigma'}\cdot \vec{S} + DS_{z}^{2}-BS_z,  \label{hk}
\end{equation}
where $c_{k\tau \sigma }^{\dagger }$ creates a conduction electron with point-group symmetry $\tau= \pm 1$ 
(channel index), spin $\sigma $ and remaining quantum numbers $k$. 
The first term describes the substrate conduction bands, the second term is the Kondo exchange interaction 
between conduction electrons and the localized spin $\vec{S}$ with exchange couplings $J_{K}^{\tau}$, $DS_{z}^{2}$ is the single-ion uniaxial magnetic anisotropy, $B$ is the magnetic field and
$\vec{\sigma}$ is the vector of Pauli matrices. 
The simplest case is when the two channels are equivalent: both are degenerate ($\varepsilon_{k \tau} = \varepsilon_k$) and the 
localized spin is equally coupled to them ($J_{K}^{\tau} = J_K$).

When intermediate valence of the magnetic impurity is included, the model is the 2-channel spin-1 Anderson 
model with anisotropy (2CS1AMA). 
It can be written in the form used first for Ni compounds with holes in the $xz$ and $yz$ 
orbitals~\cite{blesio19,mohr20,blesio23}. 
Extension to other cases, for example, FePc on Au(111)~\cite{zitko21} are straightforward. 
Neglecting the pair-hopping term~\cite{blesio19}, which is irrelevant in all cases considered so far as the 
intra-orbital Coulomb $U$ repulsion is considerably larger that the inter-orbital $U'$ one, 
the Anderson Hamiltonian is 
\begin{eqnarray}
H  & = & \sum_{k\tau \sigma }\varepsilon _{k}c_{k\tau \sigma }^{\dagger }c_{k\tau \sigma } +\sum_{k\tau \sigma }
\left( v_\tau {c}_{k\tau \sigma }^{\dagger }{d}_{\tau\sigma }+\mathrm{H.c.}\right) +   \nonumber \\
& + & \sum_{\tau \sigma }\epsilon d_{\tau \sigma }^{\dagger }d_{\tau \sigma}+\sum_{\tau }U n_{\tau \uparrow }n_{\tau \downarrow } 
+ U' n_{xz}n_{yz}-J_{H}{\vec{S}}_{xz}\cdot {\vec{S}}_{yz}+DS_{z}^{2}-BS_z,
\label{ha}
\end{eqnarray}
where $d_{\tau \sigma }^{\dagger }$ ($c^\dagger_{k \tau\sigma}$) creates a hole with energy $\epsilon$ ($\varepsilon_k$) in 
the $d$ orbital $\tau$ (conduction band $\tau$ with momentum $k$), with $\tau = xz, yz$. 
$n_{\tau \sigma }=d_{\tau \sigma}^{\dagger }d_{\tau \sigma }$ and $n_{\tau }=\sum_{\sigma }n_{\tau \sigma }$. 
$v_\tau$ is the tunneling or hybridization amplitude between impurity and conduction states 
(assumed to be independent of $k$), while $J_H$ is the strong Hund ferromagnetic exchange between the spins  
${\vec{S}}_{\tau}$ of both orbitals responsible for the total spin ${\vec{S}}={\vec{S}}_{xz}+{\vec{S}}_{yz}$ one of the impurity. The projection of ${\vec{S}}$ 
in the direction of anisotropy is denoted as $S_z$.

For small hybridization and when the two-particle configuration dominates, the model (\ref{ha}) reduces to the 
A2CS1KM~\cite{blesio19}.

\section{Main properties of the two-channel spin-1 models}

\label{res} 

\subsection{Two equivalent channels at zero magnetic field}

\label{simple} 

In the simplest case, without spin or channel anisotropy 
($B=0$, $J_{K}^{\tau} = J_K$ independent of $\tau$), it has been found that the TQPT in the A2CS1KM occurs at an anisotropy $D_c \sim 2.5\;T_K$, 
where $T_K$ is the Kondo temperature for $D=0$~\cite{blesio19}.  
The evolution of the spectral density for localized states $\rho(\omega)$ with $D$ for constant  
$T_K$ is shown in Fig.  11 of Ref.~\cite{blesio19}, where an abrupt change at the Fermi level signals the critical 
value $D_c$

\begin{figure}[hb]
\begin{center}
\includegraphics[width=0.8\textwidth]{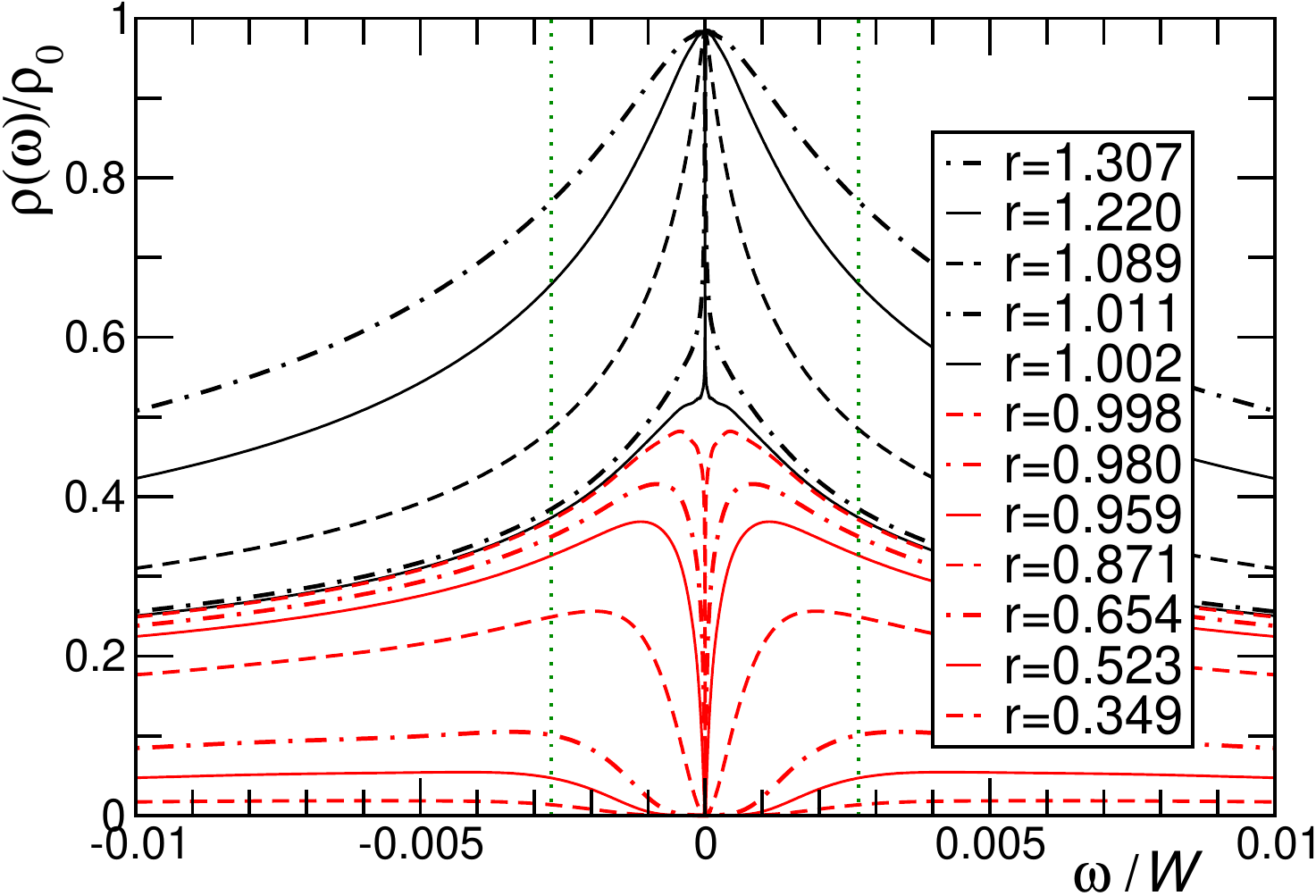}
\end{center}
\caption{(Color online) Spectral density of localized electrons of the A2CS1KM as a function of energy for 
several values of $r=J_K/J_{Kc}$, where $J_{Kc}$  is the value of $J_K$ at the TQPT. 
$\rho_0$ is given by the conventional Friedel sum rule with vanishing Luttinger integral~\cite{zitko21}. 
Vertical dotted lines are at $\omega= \pm D$. $\omega = 0$ corresponds to the Fermi level.}
\label{totra}
\end{figure}

In Fig. \ref{totra} we represent $\rho(\omega)$ for different Kondo exchange couplings $J_K$, keeping $D = 0.0027\;W$ constant, 
where $W$ is the half-bandwidth of the conduction bands, taken as the energy unit. 
$r \equiv J_K/J_{Kc}$, where $J_{Kc}$ is the critical Kondo coupling for the given $D$ ($J_{K}$ such that 
$D \simeq 2.5\; T_K$).
The numerical calculations were performed with the Ljubljana code of the NRG~\cite{zitko09,nrglj}. 
We assume  flat conduction bands extending from $-W$ to $W$ for both symmetries. 
 
For $r$ significantly larger than 1, the spectral density of localized states is similar to that of a conventional Kondo peak. 
For $r$ slightly larger than 1, $\rho(\omega)$ has the form of a narrow peak mounted on a broad peak. 
The latter, at the Fermi level, has around half the magnitude expected for the usual compensated Kondo model and therefore, 
it is similar to the spectral density expected for the spin-1/2 two-channel Kondo model. 
However, as long as $r > 1$, the system satisfies the conventional Friedel-Langreth sum rule (the small deviation in the figure 
is due to numerical errors of the NRG) and $\rho(0)$ has its maximum possible value. 
However, as soon as $r<1$, $\rho(0)$ jumps to 0. This is due to a jump in the Luttinger integral from 0 to $\pi/2$ at 
$r=1$~\cite{blesio18,zitko21}, as discussed below. If $r$ is decreased further, the shape of the spectral density tends to 
be rectangular, with two jumps at $-D$ and $D$, typical of inelastic scattering when only one channel is 
present~\cite{zitko08,zitko10,parks10,cornaglia11}. These steps are overbroadened in Fig.~\ref{totra} due to technical reasons 
that limit the resolution of the NRG at large energies~\cite{bulla08}.

A detailed description of what happens with the Friedel-Langreth sum rule and the Luttinger integrals in the more general case,
which includes intermediate valence, different channels, and a magnetic field has been explained in detail in the Supplemental 
Material of Ref.~\cite{zitko21}. To avoid including many technical details, we outline the main facts for the simpler
case of equivalent conduction channels and zero magnetic field in the two-orbital Anderson model, 
for which the Luttinger integral $I_L$ does not depend on the spin and channel quantum numbers.

Using conservation laws, it can be shown that the spectral function of the localized states for each orbital and spin, at the 
Fermi level and $T=0,$ is given by 
\begin{equation}
\rho_{\tau\sigma}(\omega=0)=\frac{1}{\pi \Delta_{\tau}} \sin ^{2}(\delta_{\tau\sigma}),
\label{fsr}
\end{equation}
where $\Delta_{\tau}= \pi v_{\tau}^2 \rho_c$, with  $\rho_c$ the density
of conduction states assumed independent of energy. The phase shift suffered by the conduction electrons at the Fermi level 
due to the presence of the impurity is
\begin{equation}
\delta _{\tau\sigma}= \pi \langle n_{\tau \sigma} \rangle -I_{L}^{\tau \sigma}.
\label{delta}
\end{equation}
The Luttinger integral $I_L$ (independent of orbital and spin indices in the simplest case) is defined as
\begin{equation}
I_{L}^{\tau \sigma} ={\rm Im} \int_{- \infty}^{0} d \omega G^d_{\tau \sigma}(\omega) 
\frac{\partial \Sigma^d_{\tau \sigma}(\omega)}{\partial \omega},  
\label{il}
\end{equation}
where $G^d_{\tau \sigma}(\omega)$ is the impurity Green function for orbital $\tau$ and spin $\sigma$, and 
$\Sigma^d_{\tau \sigma}(\omega)$ is the corresponding self energy.

For a long time, $I_L$ has been assumed to vanish for a Fermi liquid, based on perturbation calculations starting 
from a non-interacting electronic system~\cite{luttinger60b,langreth66,yoshimori82}. However, rather 
recently~\cite{curtin18,nishikawa18} it has been found that this is not always the case for local Fermi liquids. 
A topological interpretation of $I_L$ was provided for extended systems in Ref.~\cite{seki17} and extended to the 
impurity case in Ref.~\cite{blesio18}. 
Specifically, defining the ratio of non-interacting 
and interacting Green functions
\begin{equation}
 D_{\tau\sigma}(z) = \frac{G^{d(0)}_{\tau\sigma}(z)}{G^d_{\tau\sigma}(z)},
\end{equation}
it has been shown that $I_{L}^{\tau \sigma}$ is given by 
the sum of the winding number of $D_{\tau\sigma}(z)$ 
around a circuit that surrounds the negative frequency axis plus half the corresponding winding number for a circuit that surrounds the origin (see 
the supplemental material of Refs. \cite{blesio18}
and \cite{zitko21}).

An explicit calculation has shown that in the \textquotedblleft non-Landau \textquotedblright phase, for large $D/T_K$, 
$I_{L}^{\tau \sigma}=\pi/2$~\cite{zitko21}. 
In the Kondo limit, $\langle n_{\tau \sigma} \rangle=1/2$ and Eq. (\ref{fsr}) gives $\rho_{\tau\sigma}(0)=0$
in this phase, whereas in the conventional Fermi liquid phase, with 
$I_L=0$, the spectral density at the Fermi level has its  maximum possible value
$\rho_{\tau\sigma}(0)=1/(\pi \Delta)$. This explains the jump observed in Fig. \ref{totra} at the TQPT.

Previously a non-trivial value of $I_L$ has been found in underscreened 
one-channel spin-1 models, for which the system is a singular Fermi liquid~\cite{logan09}.

We note that simpler alternative approaches to NRG 
fail to capture the QPT. While the non-crossing approximation can describe non-Fermi liquid behavior quite well \cite{dinapoli14} it does not capture correctly the ``non-Landau'' phase \cite{blesio19} or in general 
phases where the isolated impurity has a non-degenerate ground state \cite{roura10}). The slave-boson mean-field approximation \cite{aligia22} reduces the system to a non-interacting one, thus always resulting in a conventional Fermi liquid. 
For single-channel systems, perturbation theory up to third order in the exchange interaction $J_K$ has successfully explained certain experiments involving spins $S>1/2$ for $D \gg T_K$ \cite{Ternes}. However, it becomes unreliable for larger $T_K$, and extending this approach to two-channel systems is cumbersome.

In Ref.~\cite{blesio18}, the conductance through a two-channel spin-1 impurity was calculated 
using the Anderson Hamiltonian Eq. (\ref{ha}). Curiously at the quantum critical point and in a more extended 
critical region at finite temperatures, the spectral density and the conductance resemble those of a spin-1/2 
two-channel Kondo model, which is a non-Fermi liquid. 
The value of the conductance at zero voltage and that of the spectral density at zero frequency are both half 
those expected in the Kondo regime for a conventional Fermi liquid.
This is further confirmed by an analysis of the eigenvalues from the NRG flow, which we briefly comment on below. 
For more detailed information, interested readers can refer to the supplemental material in Ref.~\cite{blesio18}.

Increasing the length of the Wilson chain $N$ corresponds to lower temperatures. 
For values of the anisotropy $D$ close to the critical one $D_c$ and small $N$, the system is in
the regime of the spin-1/2 two-channel Kondo model. The NRG energy flow exhibits an extended plateau, 
without even-odd alternation in the eigenvalues, signaling the existence of an unstable fixed point at intermediate temperatures. For larger $N$ (smaller temperatures), the system enters in the regime of 
one of the Fermi liquids depending on the sign 
of $D-D_c$ (topological for positive $D-D_c$). The flow 
is typical of the Fermi liquid, with even-odd alternation. At the large $N$ fixed points there is a simple relation between the eigenvalues of both cases: 
$E_{N+1}(D>D_c)=E_{N}(D<D_c)$. 
This can be understood considering that, in the strong-coupling fixed point and for $D>D_c$ the impurity is decoupled, 
while for  $D<D_c$ it is strongly coupled to the first site of the conduction chain, effectively removing it 
from the non-interacting chain.

Increasing the length of the Wilson chain $N$ corresponds to lower temperatures. For values of the anisotropy $D$
close to the critical value $D_c$ and small $N$, the system is in the regime of the spin-$1/2$ 
two-channel Kondo model. The NRG energy flow exhibits an extended plateau, with no even-odd alternation in the 
eigenvalues, indicating the presence of an unstable fixed point at intermediate temperatures. For larger $N$ 
(or lower temperatures), the system enters the regime of one of the Fermi liquids, depending on the sign of 
$D - D_c$ (topological for $D - D_c > 0$). 
The flow is characteristic of a Fermi liquid, with even-odd alternation. At the large $N$ fixed points, 
there is a simple relationship between the eigenvalues of the two cases:
$E_{N+1}(D > D_c) = E_{N}(D < D_c)$.
This can be understood by noting that, at the strong-coupling fixed point, for $D > D_c$, 
the impurity is decoupled, while for $D < D_c$, it is strongly coupled to the first site of the conduction chain, effectively removing it from the non-interacting chain.

\subsection{Effect of channel anisotropy}

\label{difcha} 

As stated above, for equivalent channels ($J_{K}^{-1}=J_{K}^{1}$) the critical value of $D$ at the transition is 
$D_c \sim 2.5\;T_K$. From results with only one 
channel \cite{parks10,cornaglia11} one knows that
$D_c \rightarrow 0$ for $J_{K}^{-1}¨ \rightarrow 0$.

We define the Kondo temperature $T_K^{(\tau)}$ as the temperature for which the contribution of
channel $\tau$ to the zero-bias conductance for $D=B=0$ falls to half of its zero-temperature value \cite{blesio18}.

\begin{figure}[ht]
\begin{center}
\includegraphics[width=0.7\textwidth]{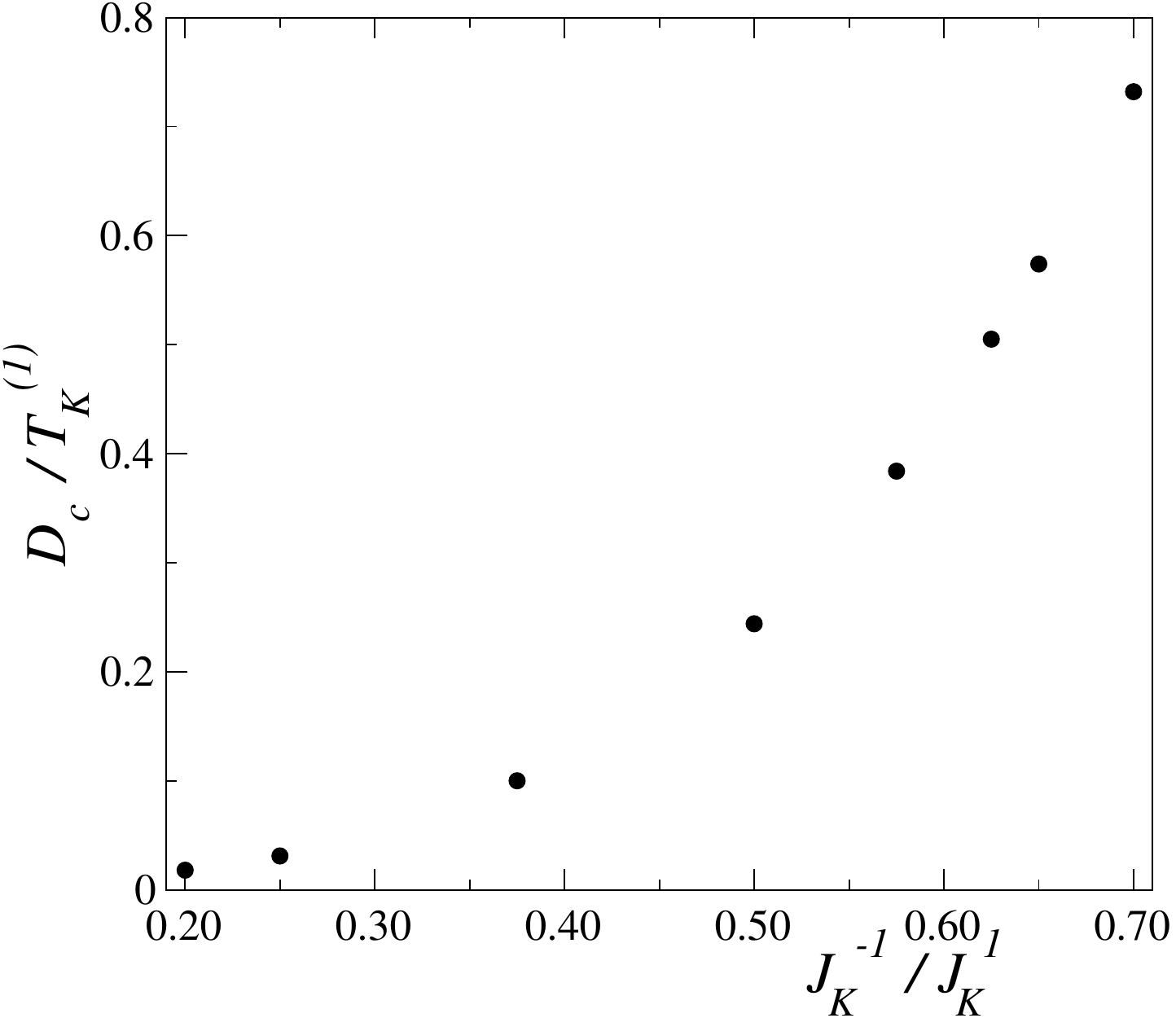}
\caption{Ratio of the critical anisotropy $D_c$ over 
the higher
Kondo temperature, varying 
the smaller exchange coupling $J_{K}^{-1}$ with
constant larger coupling $J_{K}^{1}=0.4$.}
\label{dcj}
\end{center}
\end{figure}

Clearly, the ratio $D_c/T_K^{(1)}$ should decrease 
as the coupling 
ratio $J_{K}^{-1}/J_{K}^{1}$ decreases. 
In Fig. \ref{dcj} 
we display the relation between both quantities near
$J_{K}^{-1}/J_{K}^{1}=0.5$ which 
appears to be relevant ratio for iron phthalocyanine on Au(111) (see Section \ref{fepc}). Decreasing
$J_{K}^{-1}/J_{K}^{1}$ from 1 to 0.5, $D_c/T_K^{(1)}$
has decreased by an order of magnitude, making it 
feasible to reach the ``non-Landau'' phase in the 
system with a moderate anisotropy.
We note that the highest Kondo temperature $T_K^{(1)}$
increases slightly with decreasing $J_{K}^{-1}$ 
($T_K^{(1)} \sim 0.021 \pm 0.03$ in the range of 
the figure).
Instead, $T_{K}^{-1}$ decreases strongly \cite{zitko21}.

\subsection{Effect of magnetic field with channel anisotropy}

\label{mag} 

The strong decrease in $D_c$ with channel anisotropy makes it experimentally feasible to observe the 
effects of a magnetic field at moderate intensities, as we will discuss in Section~\ref{comparison}. 

However, the inclusion of a magnetic field complicates the theory, as it breaks both spin and channel 
symmetries. When channels and spins are equivalent, Eqs. (\ref{fsr}) and (\ref{delta}) are derived using a 
single conservation law. If one of these symmetries is broken, there are two conservation laws, 
corresponding to two independent Luttinger integrals $I_{L}^{\tau \sigma}$ [see Eq. (\ref{il})]. 
When both symmetries are broken one has three conservation laws (for total number of particles, spin and 
channel pseudospin) leading for three topological numbers for four Luttinger integrals. 
A detailed analysis (see supplemental material of Ref. \cite{zitko21}) shows that as soon as a
non-zero magnetic field $B$ is introduced, the Luttinger integrals lose their topological nature and 
become continuous variables. Therefore the system becomes a conventional Fermi liquid. However, 
all physical properties are continuous, in particular the conductance.

\begin{figure} [ht]
\begin{center}
\includegraphics[width=0.8\textwidth]{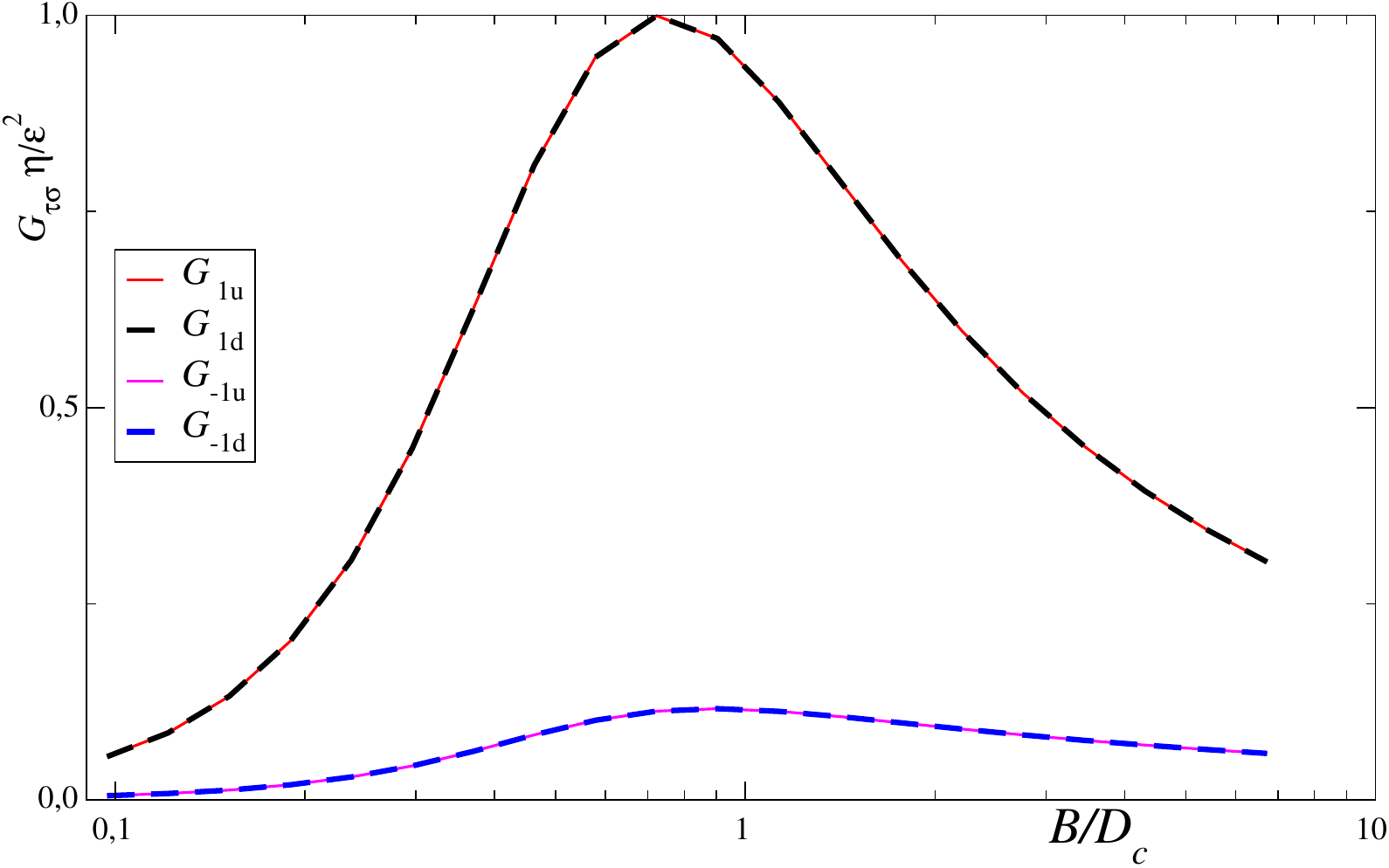}
\caption{Contribution to the conductance for each channel 
and spin in the Anderson model as a function of magnetic field for $D/D_c=1.67$.}
\label{gb}
\end{center}
\end{figure}

In Fig. \ref{gb} we show the contribution to the conductance at zero temperature for each channel and spin 
$G_{\tau \sigma}$ of the Anderson model Eq. (\ref{ha}) for parameters 
$U=0.4$, $U^\prime=0$. $\epsilon=-U/2$, $J_H=0.1$, $\Delta_1=0.06$, and $\Delta_{-1}=0.0345$. Since these quantities are related with the corresponding phase shifts

\begin{equation}
G_{\tau\sigma}(\omega=0)=\frac{e^2}{h} \sin ^{2}(\delta_{\tau\sigma}),
\label{gtau}
\end{equation}
they also give information of the spectral densities at the Fermi level [see Eq. (\ref{fsr})].

For small $B$, all $G_{\tau \sigma}$ are very small,
as expected from the proximity of the system to the 
``non-Landau'' phase for $B=0$ and $D>D_c$. As the 
magnetic field increases, the contributions of the 
more strongly coupled channel $G_{1 \sigma}$, which are the same for both spins, increase and for $B \sim 0.7 D_c$ they reach 
the maximum value, characteristic of the conventional
Fermi liquid.  For larger magnetic field $G_{1 \sigma}$
decrease. This is also expected in the conventional Kondo
regime, because the Kondo peak splits between spins up
and down as $B$ becomes larger than the Kondo temperature
$T_K^{(1)}$.

In contrast, the contributions of the more weakly coupled
channels $G_{-1 \sigma}$ remain very small, although
qualitatively the dependence with magnetic field is similar.

\begin{figure}[hb]
\begin{center}
\includegraphics[width=0.95\textwidth]{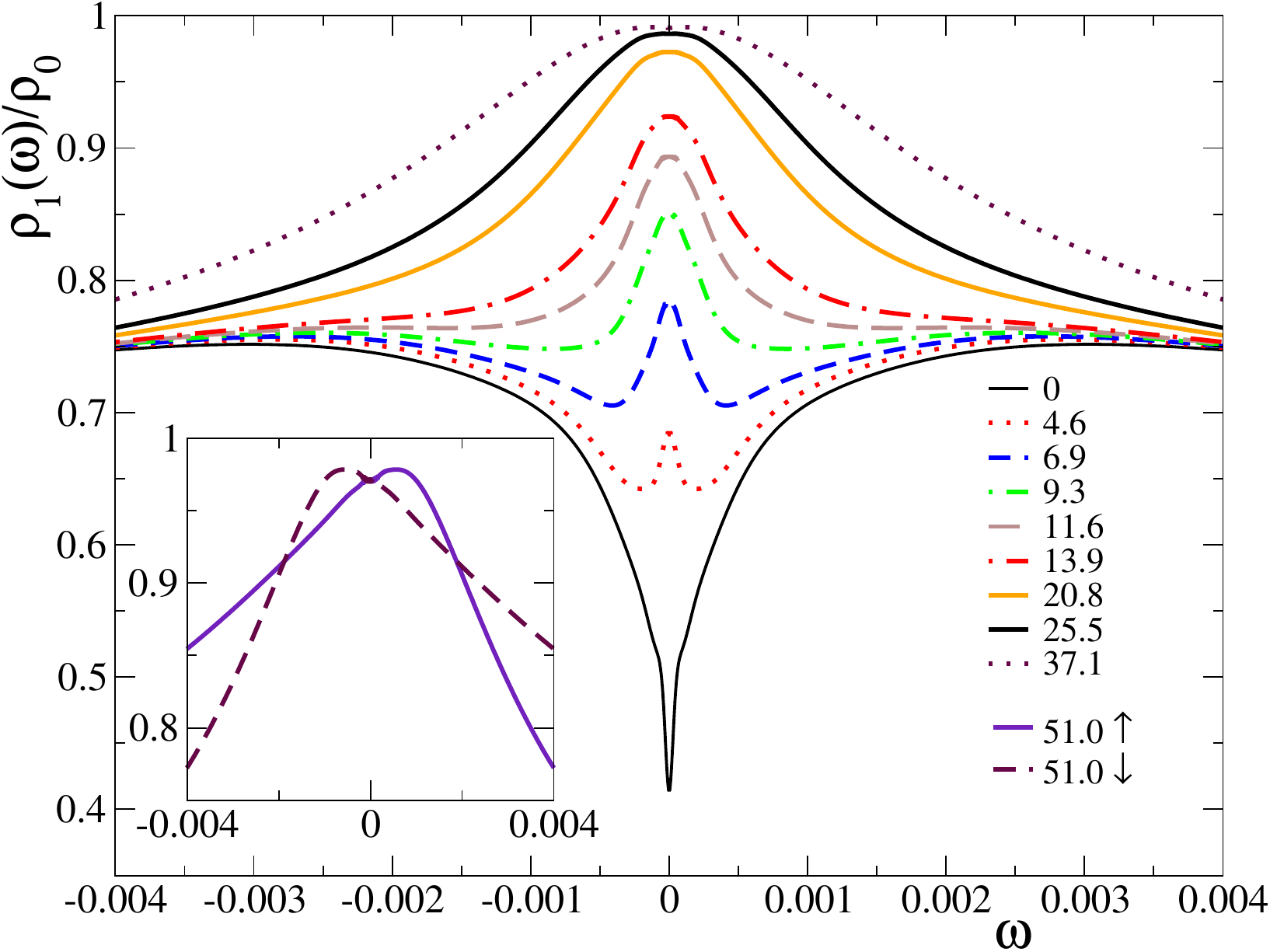}
\end{center}
\caption{(Color online) Spectral density of localized electrons of channel 1 in  the A2CS1KM as a function of energy for 
several values of magnetic field in units of $10^{-4}$ for 
$J_K^1=0.44$, $J_K^{-1}=0.22$, and $D=0.01$. Inset: spectral density resolved in spin, showing the splitting of the Kondo resonance for a large magnetic field.}
\label{rhob}
\end{figure}

In Fig. \ref{rhob} we represent the spectral density 
$\rho_1(\omega)=\rho_{1 \uparrow}(\omega)
+\rho_{1 \downarrow}(\omega)$
of channel 1 (the most strongly coupled to the conduction
band) as a function of magnetic field for parameters
which correspond to the previous Anderson model through
a Schrieffer-Wolf transformation, and an anisotropy slightly
larger than the critical one $D_c=0.0095$. Therefore, for $B=0$, $\rho_{1 \sigma} (\omega)$ has a narrow dip 
of width $\sim 10^{-3}$ similar to those displayed in
Fig. \ref{totra} for $D$ slightly larger than $D_c$.
However while in that case, increasing the exchange couplings 
leads to an abrupt TQPT, in the present case, the spectral density changes continuously. For a magnetic field of the order of the width of the dip at $B=0$, the dip disappears 
and leaves its place to a narrow peak that broadens. For larger magnetic fields, the splitting of the peaks 
for both spins begin to be important and the total spectral density at the Fermi level $\omega=0$ decreases again. Note that  $T_K^{(1)} \sim 0.034$ and thus, a complete 
splitting requires larger fields. This situation is similar to that observed in Mn phthalocyanine (see Section \ref{mnpc}).

\section{Equivalence with two-impurity systems}
 
\label{equiv}

In this Section we that the two-channel spin-1 Anderson model with single-ion anisotropy [Eq.~(\ref{ha})] 
can be exactly mapped to a system of two $S=1/2$ impurities coupled through an anisotropic exchange interaction between them. The 
Hamiltonian is 
 
\begin{eqnarray}
H_{2I}  & = & \sum_{k\alpha \sigma }\varepsilon _{k}c_{k\alpha \sigma }^{\dagger }c_{k\alpha \sigma } +\sum_{k\alpha \sigma }
\left( v {c}_{k\alpha \sigma }^{\dagger }{d}_{\alpha\sigma }+\mathrm{H.c.}\right) +   \nonumber \\
& + & \sum_{\alpha \sigma }\epsilon^{\rm 2I} d_{\alpha \sigma }^{\dagger }d_{\alpha \sigma}+\sum_{\alpha }U^{\rm 2I} n_{\alpha \uparrow }n_{\alpha \downarrow } 
+ J_z s^z_1 s^z_2 + \frac{J_\perp}{2}\left(s^+_1 s^-_2 + s^-_1 s^+_2\right).
\label{h2i}
\end{eqnarray}
For simplicity we consider two equivalent channels, and 
$\alpha = 1, 2$ refers to both impurities.

Changing the basis of the four impurity states from the individual spin projections $|s^z_1 s^z_2 \rangle$ to the total spin and projection $|SM\rangle$ with $S=0,1$, $M=-1,0,1$, and denoting by $E_{SM}$ the energies of these states in the new basis,  one obtains an effective Hund coupling $J_H=E_{00}-E_{10}=-J_\perp$ and     
an exchange anisotropy  directly related with the single-ion anisotropy of 
the spin-1 model through the relation 
$D =E_{10}-E_{10}= \frac{1}{2} (J_z - J_\perp)$.

It is interesting to note that the properties of the system are invariant under a rotation of one of the impurity spins (for example $s_2$) in $\pi$ around the $z$ axis (or equivalently a change of sign of the states with $s^z_2=-1/2$). This transformation changes the sign 
of $J_\perp$ and interchanges the states $|00\rangle$
and $|10\rangle$. In particular, the isotropic two-impurity model with $J_\perp=J_z>0$ is mapped into our A2CS1KM model
with $J_H=D=J_z$, for which one expects a topological
phase for low or moderate $T_K$.

Therefore, the two-impurity system will undergo a topological quantum phase transition depending on the exchange anisotropy. 
In fact, one of the first systems in which the existence of a non-trivial Fermi liquid was detected was the two-impurity system in 
Ref.~\cite{curtin18}. 
This significantly broadens the range of impurity systems in which the TQPT theory could be useful to interpret scanning-tunneling 
spectroscopy measurements. 

Other equivalent models can be constructed interchanging the spin degree of freedom with the pseudospin one that
takes into account the charge degrees of freedom. 
In fact, this equivalence has been used in Section III B of Ref. \cite{deleo04} to map an Anderson model similar to ours to the corresponding version with pseudospin interactions. 
However, these models do not seem to be realistic.

\section{Extension to larger spin}

\label{exte} 

For impurities involving transition metal ions, the localized spin $S$ can vary in the interval $0 \leq S \leq 5/2$.
Clearly, for $S=0$ there is no Kondo effect. The number of channels can also vary reaching up to 5 channels. 
Since treating more than two channels is computationally too expensive with the NRG, we restrict the present study to 
two channels, assuming for the moment equivalent channels.

The effects of anisotropy when only one channel is present was studied before~\cite{zitko08,zitko10,parks10,cornaglia11}.
For $S=1$, the transition occurs at $D_c = 0^+$~\cite{parks10,cornaglia11} and the Luttinger integral takes the value 
$\pi/2$ at this point~\cite{logan09}. For any single-channel $S \ge 1$ impurity, there is a competition between the underscreened 
Kondo effect~\cite{nozieres80} and the single-ion anisotropy for $D > 0$, giving rise to a complex low energy behavior that 
strongly depends on the integer or half-integer nature of $S$ and the $D=0$ Kondo temperature $T_K$ (see Table~\ref{kondos}).

For two channels and $S=1/2$, the anisotropy is irrelevant and one has a non-Fermi liquid behavior corresponding to 
the spin-1/2 two-channel Kondo model (2CKM)~\cite{mitchell12}. The $S=3/2$ case has been studied before with 
NRG~\cite{dinapoli13,dinapoli14}, and it was found that the two-channel Kondo effect also 
describes its low-energy physics as in the $S=1/2$ case. 
On the other hand, for $S=2$ we find a similar topological transition as for $S=1$. 
Therefore, the low-temperature behavior can be divided into two groups depending if the spin is integer or a half-integer. 
Table~\ref{kondos} presents a summary of the different cases.\\

\begin{table}[htbp]
\begin{center}
\begin{tabular}{|c|c|l|l|}
 \hline
    $\#$ ch.& $S$  & D &  Low energy electronic state   \\ \hline
     1 & 1/2  & irrelevant & fully compensated KE, conventional FL\\
      \hline
     1  &       integer  & $D_c =  0^+$ & two-stage KE: \\
     & & & effective (anisotropic) S=1/2 KE for $D < T_K;$ \\
             & & & quenched impurity spin for $D > T_K$ ~\cite{zitko08} \\
             \hline
1            & half-integer  & $D_c =  0^+$ & two-stage KE: quenched impurity spin for $D < T_K;$\\
      & & & effective and complex S=1/2 KE for $D > T_K$ ~\cite{zitko08}   \\
                \hline
     2 & 1/2 & irrelevant & two-channel KE: non-Fermi liquid~\cite{nozieres80} \\
     \hline
2       & 1 & $D_c > 0$ & $ D < D_c$: fully compensated KE  \\
       &  &  & $D > D_c$:  quenched impurity spin, {\it non-Landau} FL~\cite{blesio18}  \\
       \hline
2       & 3/2 & $D_c = 0^+$ & effective two-channel (anisotropic) KE~\cite{dinapoli13} \\
\hline
      2      & 2 & $D_c > 0$ & $ D < D_c$: fully compensated KE  \\
       &  &  & $D > D_c$:  quenched impurity spin, {\it non-Landau} FL~\cite{blesio18}  \\
    \hline
    \end{tabular}
\end{center}
\caption{Low energy states for a magnetic impurity of spin $S$ with $D > 0$ and coupled to $\# ch.$ equivalent 
conduction channels. KE = Kondo effect; FL = Fermi liquid. $T_K$ is the Kondo temperature for $D=0$.}
\label{kondos}
\end{table}

\begin{figure}[h]
\begin{center}
\includegraphics[width=0.9\textwidth]{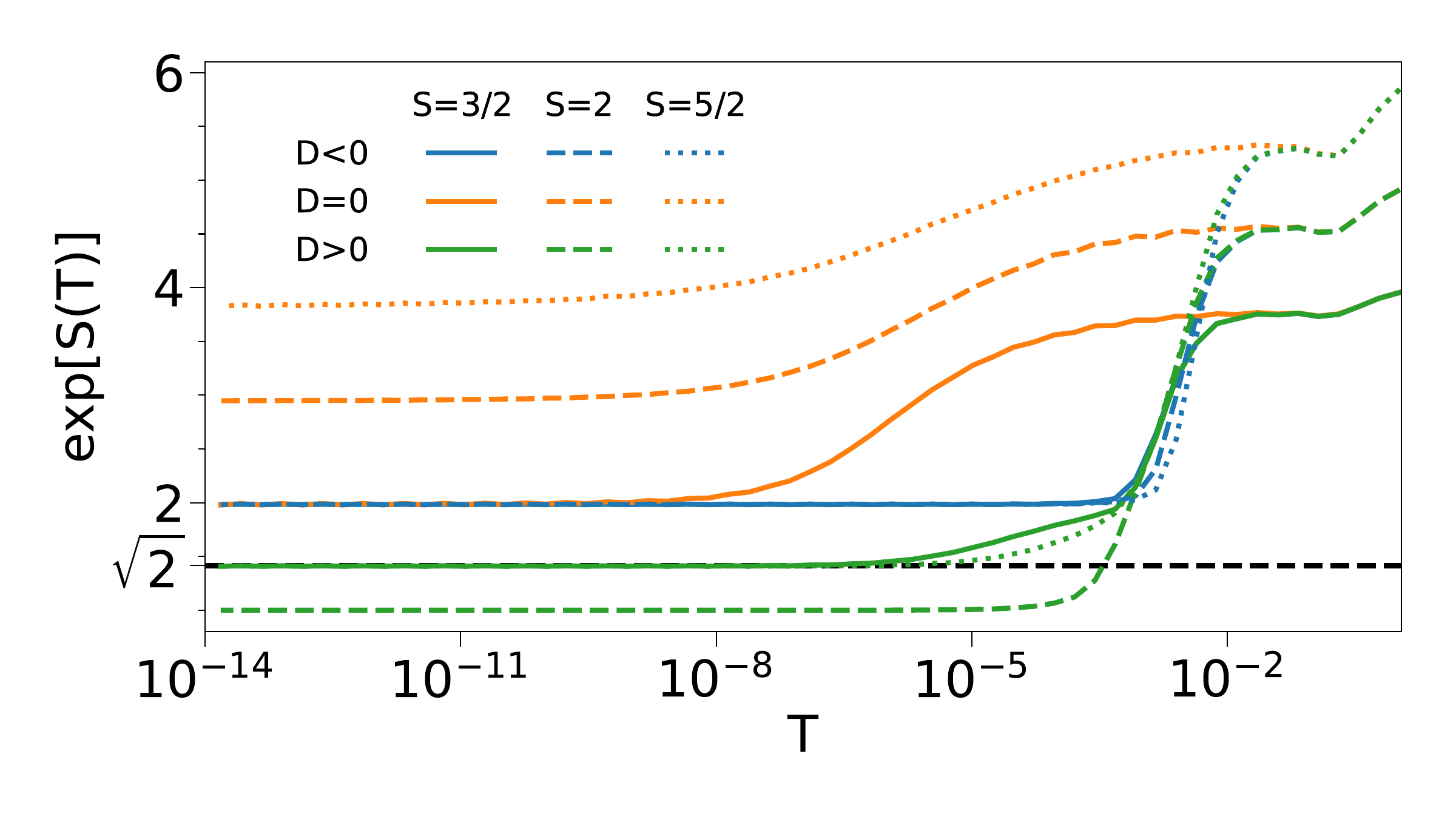}
\caption{(Color online) Entropy as a function of temperature for $J_K = 0.16$   and different values of
$S$ and $D$. $D > 0$ ($ D < 0$) corresponds to $D = 0.002$ ($-0.002$). $W$ is taken as the unit of energy.}
\label{entro}
\end{center}
\end{figure}

In Fig.~\ref{entro} we show the evolution of the entropy with temperature for different spins and anisotropy. 
For $D=0$ (orange curves), the two equivalent hybridizing channels reduce the ground state spin to 
a residual value $S^\prime=S-1$ and the entropy at zero temperature approaches ln($2S^\prime-1$). 
For negative $D$ (blue curves) the ground state is two-fold degenerate. 
The most interesting case is for positive $D$. In this case, for half-integer $S$ the low-energy behavior is 
dominated by the physics of the spin-1/2 two-channel Kondo model and the zero-temperature entropy is $\frac{1}{2}\ln(2)$. 
For integer $S$, the ground state is non-degenerate. 

\begin{figure}[h]
\begin{center}
\includegraphics[width=0.85\textwidth]{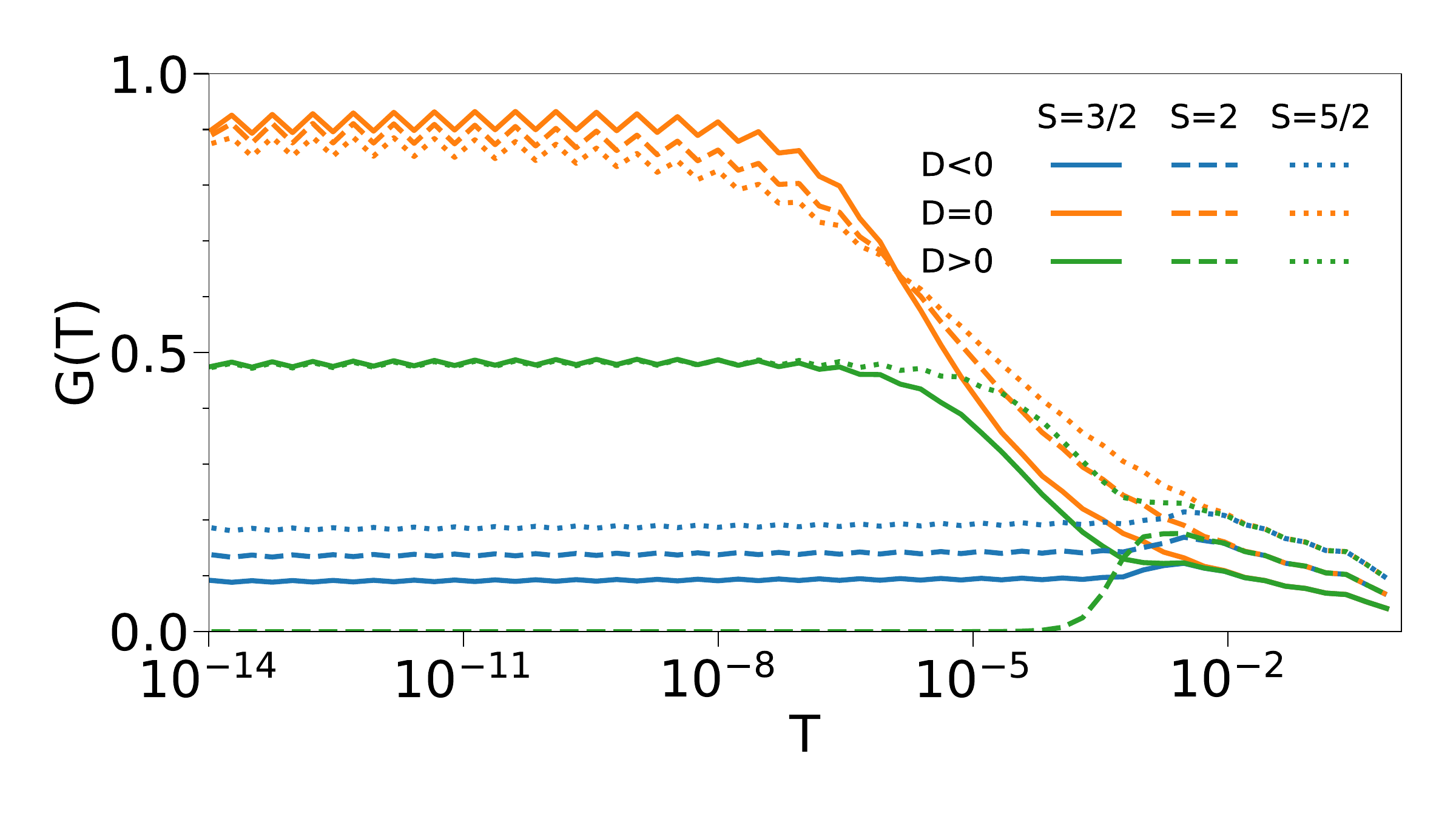}
\caption{(Color online) Conductance as a function of temperature  for $J_K = 0.16$ and different $S$ and $D$. 
$D > 0$ ($ D < 0$) corresponds to $D = 0.002$ ($-0.002$). $W$ is taken as the unit of energy.}
\label{cond}
\end{center}
\end{figure}

In Fig.~\ref{cond} we show the conductance $G(T)$ (related to the spectral density) as a function of temperature for 
different $S$ and $D$. For $D=0$ (orange curves), $G(T)$ for $T \rightarrow 0$ tends to the unitary 
value $G_0=4 e^2/h$, 
characteristic of two orbital- and spin-degenerate conduction channels. 
The case $D < 0$ is complex and is not of interest to the central objective of this article, because in such a case, 
there is no quantum phase transition. 
For $D > 0$ and half-integer $S$, the low-temperature value is $G_0/2$, characteristic of the spin-1/2 
two-channel Kondo model. 
For integer $S$, $G(0)=G_0$ (zero) if $D$ is below (above) the critical anisotropy $D_c$ of the topological 
transition. In the figure only the case $D > D_c$ is shown.

We have also analyzed what happens for the half-integer case when the couplings of the two channels 
are different in presence of a small $D > 0$. 
The ground state becomes a singlet and therefore, the entropy goes to zero for $T \rightarrow 0$. 
The contribution to the conductance of the channel with a larger coupling constant is similar to that of the 
simplest spin-1/2 1-channel Kondo model, increasing at $T_K$ and reaching $2e^2/h$ at zero temperature.
At a lower characteristic temperature the conductance of the other channel decreases with decreasing temperature,
indicating a dip in the corresponding spectral density. This seems to be the case of MnPc on Au(111). See Section \ref{mnpc}.

\section{Experimental relevance of the topological quantum phase transition}
 
\label{comparison} 

In this section, we present several systems of atoms or molecules on metallic surfaces, and we show that 
scanning-tunneling spectroscopy experiments performed on them can be explained qualitatively or semiquantitatively by 
means of the two-channel $S=1$ Kondo or Anderson models with anisotropy. 
In most of these cases, alternative explanations were proposed in the literature that contradict basic physical 
principles or are unsatisfactory. 
As the theory of the topological quantum phase transition is relatively new and it seems to be only 
captured by NRG calculations, while alternative explanations fail, it is important to show that a consistent explanation 
using new concepts and models exists. This is our focus, and not to provide an accurate fit of all the experimental 
curves mentioned below, an objective that lies beyond the scope of this paper.

\subsection{FePc on Au(111)}

\label{fepc}

\begin{figure}[h]
\begin{center}
\includegraphics[width=0.7\textwidth]{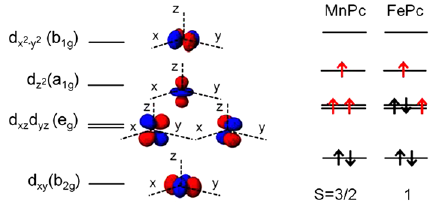}
\caption{(Color online) Electronic structure of MnPc and FePc. Reprinted with permission from Ref.~\cite{aykanat20}. 
Copyright 2020 American Chemical Society. Red arrows 
correspond to unpaired spins 1/2.}
\label{orbi}
\end{center}
\end{figure}

The system of iron phthalocyanine (FePc) on the Au(111) surface  has attracted a lot of attention during the last 15 years \cite{zitko21,gao07,bartolome10,tsukahara11,minamitani12,lobos14,fernandez15,hiraoka17,fernandez18,aykanat20,yang19b,aligia22}.
According to LDA + U calculations, the basic electronic structure of the molecule is that shown in Fig.~\ref{orbi}, 
taken from Ref.~\cite{aykanat20}. The partially filled orbitals of Fe are those of symmetry $3z^2-r^2$ with nearly one electron, 
and the degenerate so-called $\pi$ orbitals, of symmetry $xz$ and $yz$ with three electrons, resulting in a spin 1. 
This is in agreement with X-ray magnetic circular dichroism experiments~\cite{bartolome10}. 

\begin{figure}[ht] 
\begin{center}
\includegraphics[width=0.8\textwidth]{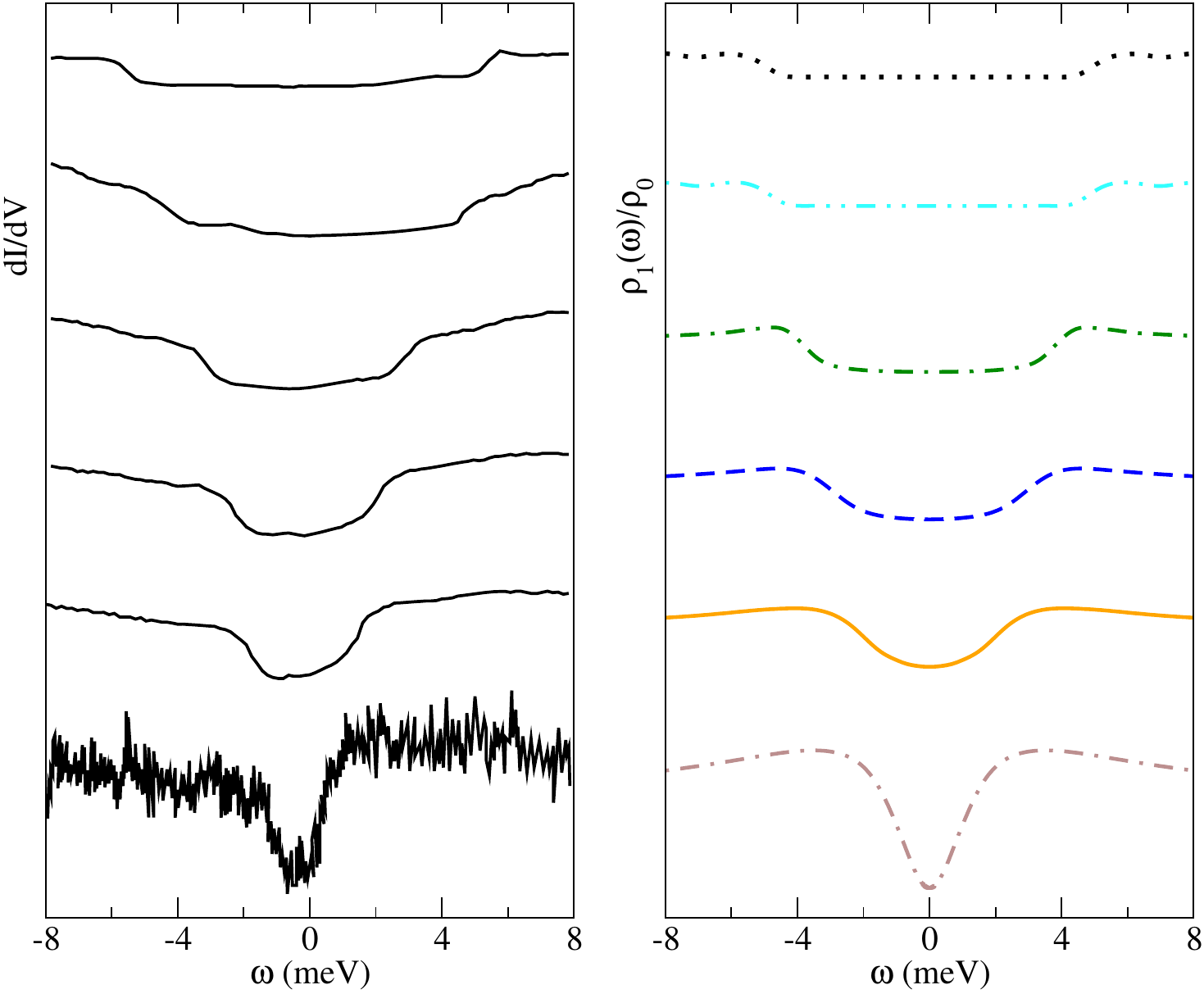} 
\caption{Left: experimental differential conductance of FePc on Au(111) as a function of voltage, taken from Fig. 2 (b) 
of Ref.~\cite{hiraoka17}; and right: theoretical spectral density of 3d $3z^2-r^2$ electrons as a function of energy, 
as the molecule is raised from the surface, taken from Supplementary Fig. 3 of Ref.~\cite{zitko21}. 
Parameters are $D=0.005$ $J_z=0.44f$, $J_a=0.22f$ in units of the band width taken as 1 eV. 
From top to bottom, the factor $f$ used is 0.1, 0.2, 0.5, 0.6, 0.7 and 0.8.}
\label{fhira}
\end{center}
\end{figure}

\begin{figure} [ht] 
\begin{center}
\includegraphics[width=0.8\textwidth]{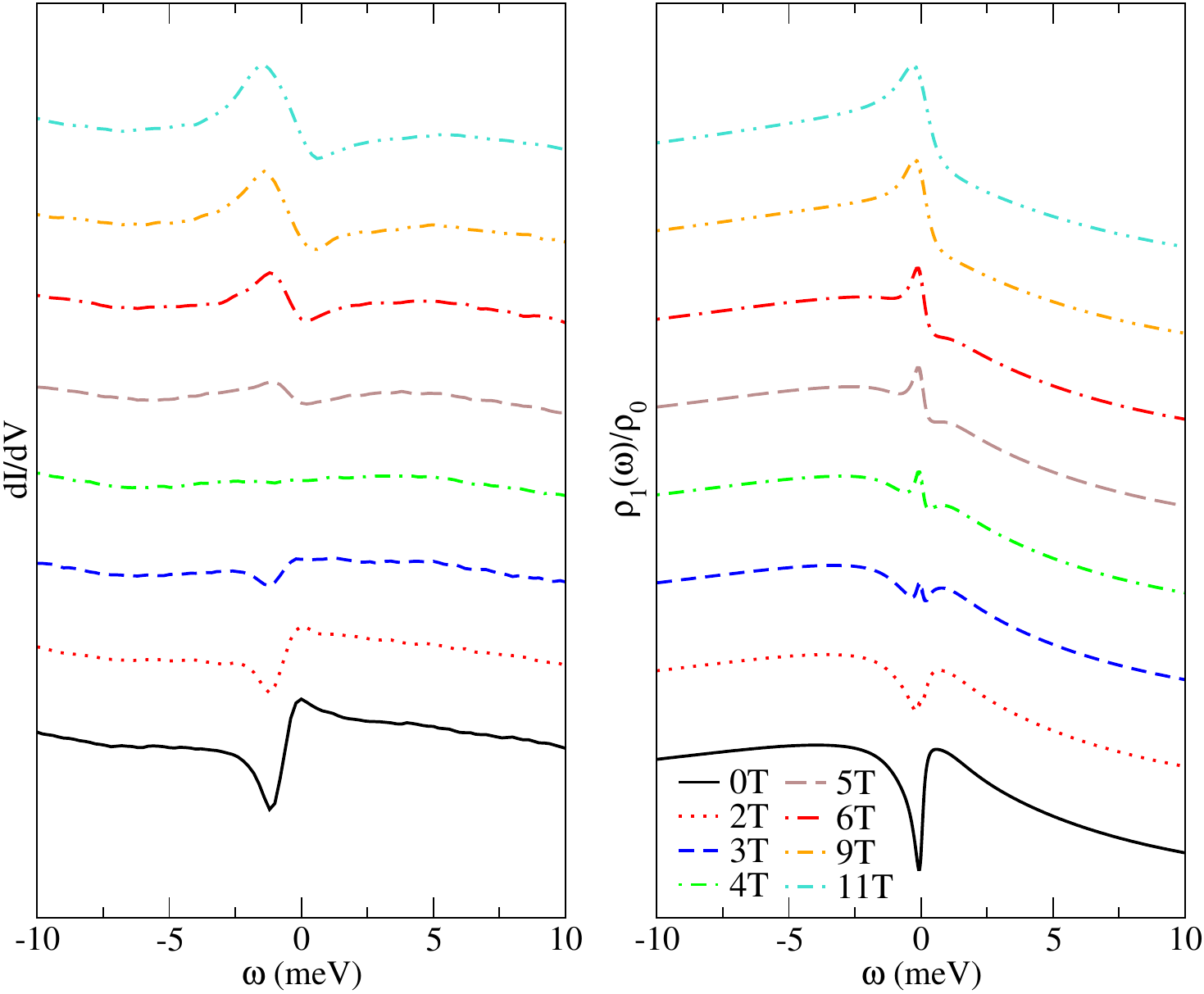} 
\caption{Experimental (left) and theoretical (right) differential conductance of FePc on Au(111) as a function of voltage 
for several values of the magnetic field. The experimental curves are taken from Fig. 2 (c) of Ref.~\cite{yang19b}, while 
the theoretical curves are taken from Supplementary Fig. 5 of Ref.~\cite{zitko21}. Parameters as in Fig.~\ref{fhira} 
with $f=1$ and asymmetry parameter $q=0.4$.}
\label{bsup}
\end{center}
\end{figure}

The differential conductance $dI/dV$ observed by scanning-tunneling spectroscopy of a single FePc molecule on Au(111) 
in the on-top position shows a narrow dip mounted on a broad peak~\cite{minamitani12,hiraoka17,yang19b} (corresponding to 
the curves at the bottom of Figs.~\ref{fhira} and~\ref{bsup}). 
This is very suggestive of the spectral density of localized electrons near the topological transition, for anisotropy 
$D$ slightly larger than the critical one~\cite{blesio18,blesio19} (See Fig. \ref{totra}). 
However, most experiments were done before the development of the theory of the TQPT and were interpreted in a different 
fashion, as a two-stage Kondo effect: the $3z^2-r^2$ orbitals hybridize strongly with the conduction electrons with the 
same symmetry, giving rise to a first-stage Kondo effect and to the broad peak in $dI/dV$ around 20 meV. 
At a lower temperature, the Kondo effect due to the $\pi$ orbitals sets in giving rise to a dip of half-width $\sim 0.6$ meV 
in $dI/dV$~\cite{minamitani12,fernandez18}. 
The fits of the spectrum suggest the following hierarchy of the different orbitals in decreasing order of 
hopping amplitude to the tip:  3d$_{3z^2-r^2}$ , conduction electrons with $3z^2-r^2$ symmetry, conduction electrons 
with $\pi$ symmetry and 3d$_\pi$~\cite{fernandez18}.

However, an experiment that can discern between both scenarios (TQPT or two-stage Kondo effect) has been made. 
Raising the FePc molecule from the surface~\cite{hiraoka17}, the hybridization amplitudes are weakened and, with them, 
the exchange interactions between localized 3d electrons and conduction electrons.  
In the two-stage scenario explained above one expects that both features, the broad peak and the Kondo dip \emph{narrow} 
since the corresponding Kondo temperatures should decrease. 
Instead, if the system is a {\it non-Landau} Fermi liquid close to the topological transition, decreasing the exchange 
interactions with respect to $D$ should \emph{broaden} the dip as the system moves away from the TQPT. 
This last scenario is what is observed experimentally (see Fig.~\ref{fhira}), giving support to the TQPT picture.

To construct the adequate model for FePc, one has to take into account the splitting of the $\pi$ orbitals as a consequence 
of the spin-orbit coupling (SOC)~\cite{aligia22}, neglected in previous treatments. The states $|\pi \sigma \rangle $ with 
one hole in the $\pi $ orbitals are (except for an irrelevant phase)
\begin{eqnarray}
|a &\uparrow &\rangle =\frac{|xz\uparrow \rangle +i|yz\uparrow \rangle }
{\sqrt{2}},|a\downarrow \rangle =\frac{|xz\downarrow \rangle -i|yz\downarrow
\rangle }{\sqrt{2}},  \nonumber \\
|b &\uparrow &\rangle =\frac{|xz\uparrow \rangle -i|yz\uparrow \rangle }
{\sqrt{2}},|b\downarrow \rangle =\frac{|xz\downarrow \rangle +i|yz\downarrow
\rangle }{\sqrt{2}}.  \label{states}
\end{eqnarray}
The $|b \sigma \rangle$ states lie above the $|a \sigma \rangle$ by an energy of the order of the SOC, estimated in 76 meV for 
Fe~\cite{fisk68}. This also leads to a significant orbital polarization which was observed~\cite{bartolome10}, and to an 
anisotropy $D \sim 5$ meV.

Therefore, an appropriate model to describe the system is the anisotropic two-channel spin-1  Kondo model (A2CS1KM), 
one channel with strong exchange coupling $J_z$ for the $3z^2-r^2$ electrons and another channel of the $a$ states $J_a$.
The effect of raising the molecule is incorporated in our model by reducing both $J_z$ and $J_a$ by the same factor $f$. 
In this way, the experiments can be semiquantitatively explained~\cite{zitko21}, as shown in Fig.~\ref{fhira}.

In Ref.~\cite{yang19b} the dependence of $G = dI/dV$ with temperature and magnetic field has been measured. 
The results have been also explained using the A2CS1KM~\cite{zitko21}. As the temperature is raised, the dip is reduced and 
disappears at $\sim 10$ K. The dependence with the magnetic field shown in Fig.~\ref{bsup} is striking: 
the narrow dip is converted into a narrow peak as the magnetic field is increased. 
The theoretical results, taken from Ref.~\cite{zitko21}, reproduce semiquantitatively the experimental data. 
To take into account the asymmetry of the shape, we assume that the STM tip senses mainly the localized electrons of symmetry 
$\tau=3z^2-r^2$ with some admixture of conduction electrons with the same symmetry weighted by the parameter 
$q$~\cite{zitko11}, which we take as $q=0.4$. The conductance given by our model $G_m$, represented 
at the right of Fig.~\ref{bsup} is therefore given by \cite{zitko21}: 
\begin{equation}
G_m(V)= -\left[(1-q^{2}){\rm Im} G_{\tau \sigma }^{d}(\omega )+2q {\rm Re} G_{\tau \sigma }^{d}(\omega)\right],
\label{gm}
\end{equation}
where $G_{\tau \sigma }^{d}(\omega )$ is the Green function of localized electrons for symmetry $\tau$ and spin $\sigma$. 

A better agreement can probably be obtained by enlarging $J_z$, which has the effect of broadening the broad peak, 
adjusting $J_a$ to broaden the dip a little bit, and by including the orbital polarization, which increases the effective 
coupling with the magnetic field by a factor of 3/2.

\subsection{MnPc on Au(111)}

\label{mnpc}

\begin{figure}[h]
\begin{center}
\includegraphics[width=0.7\textwidth]{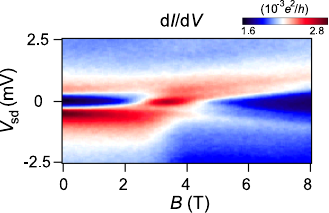}
\caption{(Color online) Differential conductance $dI/dV$ of MnPc on Au(111) as a function of magnetic field $B$ and 
bias voltage $V_{sd}$, taken from Fig. 2(d) of Ref. \cite{guo21}.}
\label{mnb}
\end{center}
\end{figure}

Curiously, in spite of having one electron less than FePc, the observed differential conductance in Mn phthalocyanine 
on Au(111)~\cite{guo21} is qualitatively very similar to that observed in the Fe system. 
There is a dip of half-width about 0.5 mV, mounted on a broad peak. 
Under the application of a magnetic field, the dip turns to a peak at $\sim 4$ Tesla, and for a larger magnetic field
the peak splits (see Fig.~\ref{mnb}). The latter behavior was not observed in the FePc system, but is expected if larger 
fields were applied in that case.

It has been suggested that the observed behavior can be explained by a singlet ground state and a triplet excited state with 
a small excitation energy~\cite{guo21}. 
However, on one hand, the Hund coupling in 3d transition-metal elements are of the order of 0.7 eV, favoring a total spin 3/2.
On the other hand, specific calculations for the singlet-triplet model with a small triplet excitation energy shows a 
$dI/dV$ that {\em decreases} slightly with increasing temperature $T$ for small $T$ (see Fig. 7 of Ref.~\cite{roura10}), in 
contrast to the experimental observations [Fig. 2 (b) of Ref. \cite{guo21}].

The calculated electronic structure for planar MnPc, is different from that in the gas phase (represented in Fig.~\ref{orbi}) 
and corresponds to the intermediate-spin quartet $^4E_g$ $[(xy)^1 (\pi)^3 (3z^2-r^2)^1]$~\cite{liao05,brumboiu14}.
This is in agreement with polarization-dependent N K-edge x-ray absorption spectra for MnPc on Au~\cite{petraki12}. 
Therefore, the difference with the electronic structure for FePc on Au(111) is that, in the Mn system, there is a hole in the 
3d$_{xy}$ which is absent in the Fe system, leading to total spin 3/2 in the Mn case.

A realistic model for MnPc on Au(111) involves therefore three channels [states with symmetry $3z^2-r^2$, $a$ of 
Eq.~(\ref{states}) and $xy$] and is almost intractable with NRG. 
In order to have a qualitative understanding, we have studied a spin 3/2, two-channel Kondo model including anisotropy, 
assuming $J_z \sim 2 J_a$ and $J_{xy}=0$.  
From the temperature dependence of the conductance, we see a first-stage Kondo effect in which the contribution of the dominant 
$3z^2-r^2$ channel is the usual one for a spin 1/2 system, saturating near $2e^2/h$ at low temperatures, without a dip. 
However, the contribution of the $a$ channel has a dip, so that the total conductance presents a narrow dip mounted on a broad peak. 
The ground state is a singlet.

The results can be qualitatively understood as follows. 
At temperatures of the order of the Kondo temperature of the dominant $3z^2-r^2$ channel (near 20 meV in the Fe system), 
the spin of that channel is screened and one is left with a spin 1 screened partially by the exchange in the $a$ channel 
and with anisotropy $D$. 
This model has been studied and, in presence of any $ D > 0$~\cite{parks10,cornaglia11}, the conductance and the spectral density 
have a dip, whose width decreases exponentially with $\sqrt{D}$, while the application of a magnetic field leads to a differential 
conductance of the same form as that shown in Fig.~\ref{mnb} of Ref.~\cite{cornaglia11}. 
If, in the effective model after screening the spin 1/2 of the dominant channel, one includes the exchange of the third channel $xy$, 
the model is precisely the A2CS1KM, and one expects a topological quantum phase transition at finite $D$. 
The parameters should be renormalized, as expected from approximate treatments of similar 3-channel models~\cite{fernandez18}.

Other impurity systems containing phthalocyanine molecules, 
such as CoPc on Au(111) \cite{wang14} and 
TiPc on Cu(110) \cite{ribeiro24} were also studied,
where the low-energy physics seems dominated by the usual Kondo effect.

\subsection{Nickelocene on Cu(100)}

\label{nc}

The system of isolated double-decker nickelocene (Nc) molecules on Cu(100) substrates have been experimentally studied 
in detail~\cite{mohr20,bachellier16,ormaza16,ormaza17,ormaza17b,verlhac19,kogler24}.
Density functional theory (DFT) calculations show that the electronic structure of Ni is basically 3d$^8$, with one hole in  
each of the nearly degenerate $\pi$ orbitals ($xz$ and $yz$), with some mixing with the 3d$^9$ 
configuration~\cite{ormaza17,mohr20}.
Therefore the appropriate model to describe the system is actually the 2CS1AMA proposed for Ni impurities in a Au chain 
doped with oxygen~\cite{blesio18,blesio19} or its integer-valent limit, the A2CS1KM~\cite{blesio19}.

 \begin{figure}[h]
\begin{center}
\includegraphics[width=0.8\textwidth]{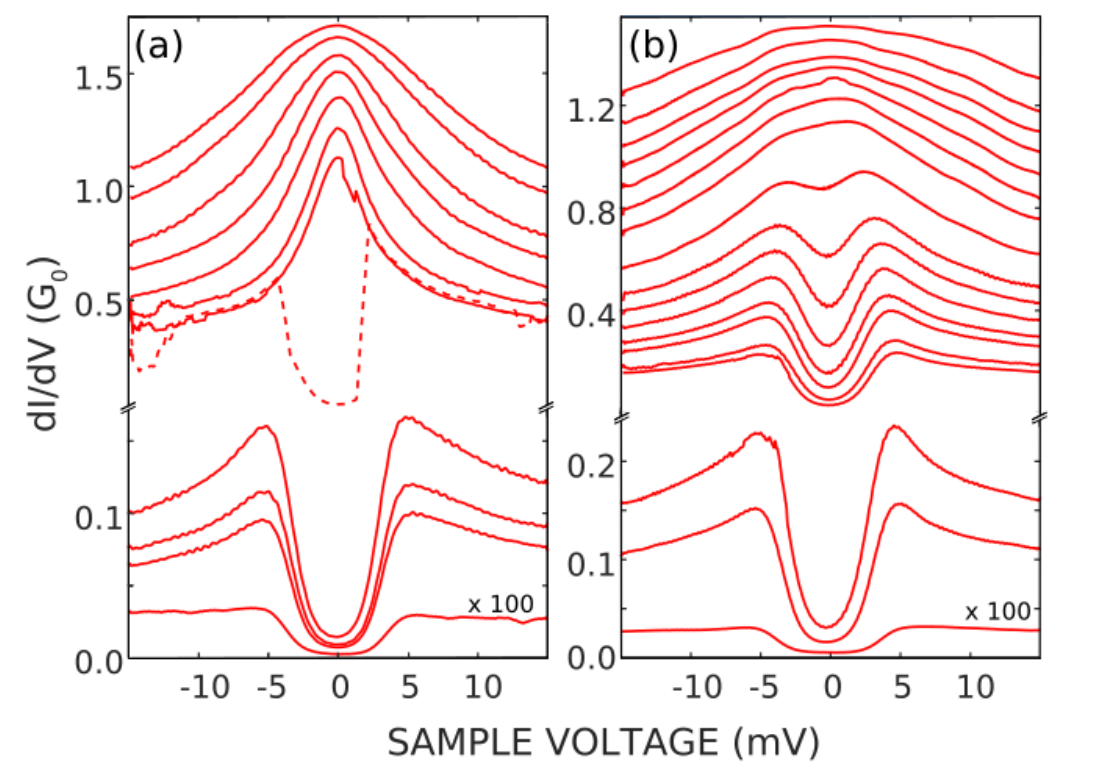}
\caption{(Color online) Differential conductance of several samples of Nc on Cu(111) 
(a) [(b)] correspond to spectra observed with frequency 2/3 [1/3], approximately. Within each panel, the higher a curve is, the 
shorter the distance from the molecule to the substrate. Taken from Fig. 5 of Ref.~\cite{mohr20}. Copyright 2020 
American Physical Society.}
\label{expnc}
\end{center}
\end{figure}

The experimentally observed spectra for the differential conductance $dI/dV$ are shown in Fig.~\ref{expnc}.  
Left and right panels correspond to spectra observed with frequency 2/3 (case A) and 1/3 (cases B), respectively. 
The curves within each panel correspond to different positions of the tip. 
As the STM tip is approached to the molecule, the hybridization between tip and molecule states increases and a 
jump from a dip to a peak near $V=0$ takes place, which is consistent with the TQPT. 

In the theory that was presented alongside the experimental observations, the occurrence of a peak or a dip has been 
tentatively ascribed to a crossover in the spin of the molecule. This transition is noted as shifting from 1/2 in 
the contact regime (STM tip near the molecule) to 1 in the tunneling regime (STM tip far from the molecule), a deduction 
based on first-principle calculations~\cite{mohr20,ormaza17}. 
However, on one hand, these calculations miss relevant dynamical correlations and, therefore, they do not properly treat 
the Kondo effect which tends to screen the spin. On the other hand, the electronic structure does not change much 
between the two regimes and, as admitted by the authors, the change in the molecular charge is actually insufficient 
to account for the large change in the spin. 

The observed spectra have many similarities with the spectral density of localized electrons in the A2CS1KM 
(see Fig.~\ref{totra}) but also important differences. 
For case A, there seems to be a first-order transition as the tip is approached to the molecule, avoiding the transition 
zone with a very narrow peak or dip, as shown in Fig.~\ref{totra}. 
For case B, the transition seems to be continuous, without a jump from a dip to a peak at zero voltage, as 
the hybridization (or exchange) between localized and conduction electrons is increased. 

The first-order transition of case A can be understood as follows:  as in FePc on Au(111), in which the molecule is 
raised when the STM approaches it~\cite{hiraoka17}, we expect that some variable $\eta$ which determines either the 
position or the shape of the molecule, modifies the hybridization $v$ of the 2CS1AMA [Eq. (\ref{ha}) assuming 
$v_\tau=v$, the same for both channels]~\cite{blesio23}, leading to a coupling of $\eta$ with our electronic model. 
In the absence of this coupling, one expects that the elastic energy is $E_e=K \eta^2/2$ (shifting the 0 of $\eta$ if necessary). 
It has been shown that the second derivative of the energy of the electronic model is strong and negative
near the TQPT~\cite{blesio23}. This means that for a soft spring (small $K$) the second derivative of the total energy
is also negative at the TQPT, leading to a first-order transition in a Maxwell construction~\cite{blesio23}.
This reasoning provides a natural explanation of the observed behavior for case A. 

\begin{figure}[ht]
\begin{center}
\includegraphics[width=0.7\textwidth]{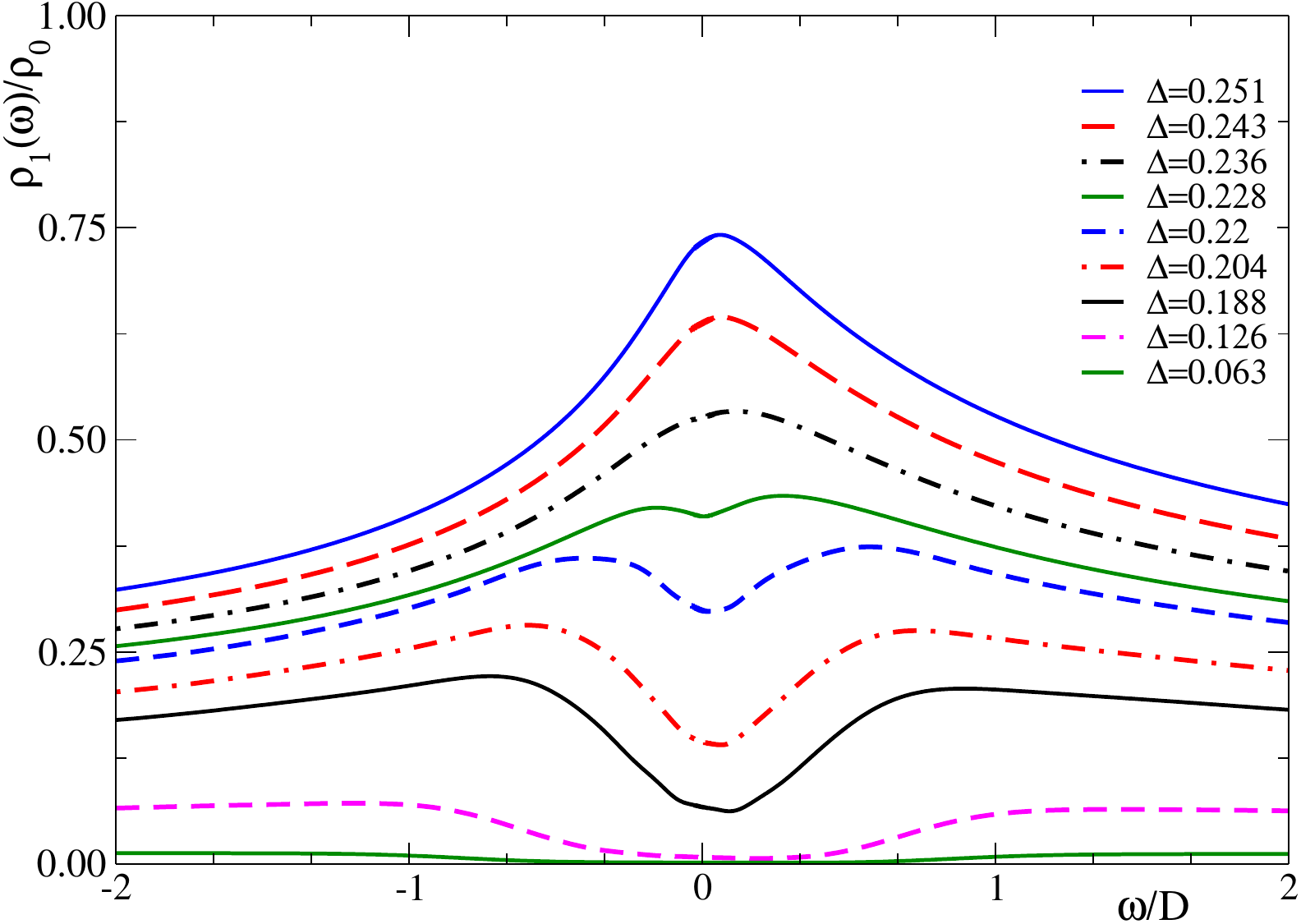}
\end{center}
\caption{(Color online) Differential conductance as a function of voltage for different values of $\Delta$. 
Other parameters are $U=3.5$, $U'=2.5,$ $J_H=0.5$, $\epsilon=-3.0$, $T=0.0005$ [see Eq.~(\ref{ha})].}
\label{figb}
\end{figure}

Case B probably corresponds to a hard spring (large $K$) and the first-order transition does not take place. 
The reason why an abrupt jump like that in Fig.~\ref{totra} is not observed is two-fold: i) the magnitude of the jump 
decreases with the degree of intermediate valence [see Eqs. (\ref{fsr}) and (\ref{delta})] and ii) finite temperature. 
This is the most relevant parameter. These effects were investigated using the 2CS1AMA described by 
Eq.~(\ref{ha})~\cite{blesio23}.
In Fig. \ref{figb} we show the evolution of the differential conductance for different values of 
$\Delta= \pi v^2 \rho_c$ where $\rho_c=1/(2W)$ is the
density of conduction electrons assumed constant in the range
$-W < \omega < W$. The half-band width $W =$ 1 eV is taken as the unit of energy.
For small $\Delta $, $dI/dV$ has a dip mounted on a broader peak, as usual. As $\Delta $ increases, 
the dip narrows, but in contrast to
the case of zero temperature, the minimum of the dip increases, and a very sharp dip like that of Fig. \ref{totra} 
is absent. 
For larger $\Delta$ the dip gradually disappears and the magnitude of $dI/dV$ near zero voltage increases. 
The overall behavior reproduces semiquantitatively the experimental results for case B shown in the right panel 
of Fig.\ref{expnc}.

\subsection{Fe Atoms on MoS$_2$/Au(111)}

\label{trish}

Trishin {\it et al.}~\cite{trishin21} studied experimentally a system consisting of an Fe atom on top of a monolayer 
of MoS$_2$ deposited in turn on a Au(111) surface. As argued below, it is very natural to expect that the system is 
described by the A2CS1KM. MoS$_2$ on Au(111) forms a Moir\'e structure, which implies strong local variations of 
the density of conduction electrons $\rho_c$. Therefore, depending on the specific position at which the Fe adatom is 
located, dramatic variations of the adimensional parameter $J'=\rho_c J_K$ (which determines the Kondo temperature) 
are expected, and one might expect to observe the TQPT as in Fig.~\ref{totra}.

\begin{figure}[tbp]
\begin{center}
\includegraphics[width=0.7\textwidth]{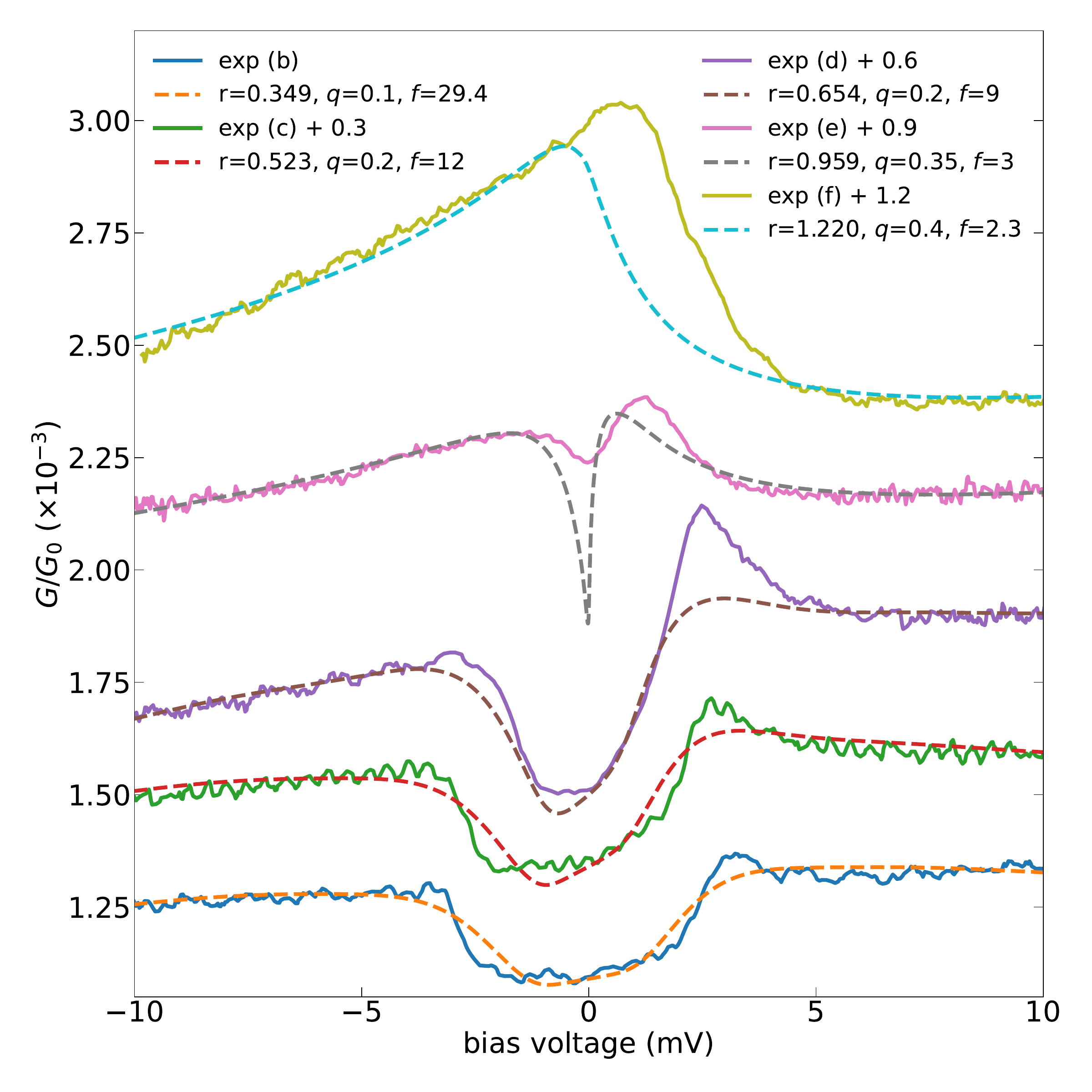}
\end{center}
\caption{(Color online) Differential conductance corresponding to the experimental curves (b) to (f) 
(full lines, each curve is labeled as exp(b),etc.) and the results for our model (dashed lines, shifted upward by the 
same magnitude as the corresponding experimental curve). Note that (a) is not included and the top figure corresponds 
to (b). See the main text for the meaning of the $r$, $q$ and $f$ fitting parameters. 
Figure taken from Ref.~ \cite{blesioaa}. Copyright 2023 American Physical Society} 
\label{moir}
\end{figure}

The validity of the A2CS1KM to describe the system can be justified as follows. DFT calculations of Fe atoms on 
free-standing MoS$_2$ indicates that the spin state of the atoms is either $S=1$~\cite{wang14b} or $S=2$~\cite{chen17}. 
However, for $S=2$, one would expect a second jump in $G(V)$ at larger $|V|$ in the regime of low $J'$, which is 
not observed experimentally [Fig. 3(h) of Ref.~\cite{trishin21}].
On the other hand, experiments and DFT calculations indicate that the Fe atoms are located in positions with 
symmetry corresponding to the point group $C_{3v}$. 
Therefore, the Fe $3d$ orbitals are split into one $A_{1}$ singlet and two $E$ doublets~\cite{morolagares18}. 
Our comparison with the experiment (shown in Fig.~\ref{moir}), indicates that the  spin 1 is formed by occupying the 
two states of an $E$ doublet (the agreement worsens when non-equivalent channels are considered).
In this case, it is clear that the spin-orbit coupling originates a hard axis anisotropy $D(S_z)^2$ with 
$D > 0$~\cite{blesioaa}. In addition, each of the degenerate orbitals of the doublet hybridizes with conduction states of 
the same symmetry~\cite{blesio18,blesio19,blesio23}. 
This reasoning naturally leads to the 2CS1AMA with degenerate channels used in Refs.~\cite{blesio18,blesio19,blesio23} 
and to the A2CS1KM in the integer valence limit.

The differential conductance $G(V)=dI/dV$ has been measured on nearly 40 different Fe positions. 
Six of them [(a) to (f)] are presented in Ref.~\cite{trishin21} and five of them [(b) to (f)] are reproduced in our Fig.~\ref{moir}. 
In Ref.~\cite{trishin21}, fits of the different spectra were done using three different approaches: 
i) perturbation theory in the exchange coupling in an $S=1$ anisotropic one-channel Kondo model for cases (a) to (d), 
ii) a Frota peak (expected for the simplest Kondo model) for case (f), and 
iii) a Lorentz peak plus Frota dip (without justification) for case (e). 
Note that the fit of the peak and the dip requires 3 parameters for each one (determining position, width and intensity) in 
addition to a linear background. Therefore the resulting good fit is not surprising~\cite{trishin21}, but it lacks a 
physical justification.

In contrast, as shown in Fig \ref{moir}, the A2CS1KM can semiquantitatively explain the data in a unified fashion.
A better agreement with the experiment can be obtained by allowing some degree of intermediate valence for cases (e) and (f),
as explained below, but we wanted to keep the number of free parameters as minimal as possible.

The numerical calculations were performed with the Ljubljana code of the NRG~\cite{zitko09,nrglj}, and were reported
previously in Ref.~\cite{blesioaa}. We assume flat conduction bands extending from $-W$ to $W$ for both symmetries. 
We take $W=1$ eV and $D=2.7$ meV. The product $J^{\prime}=\rho_c J_K$ is assumed to vary among the different cases, due to the 
Moiré modulation. The TQPT is at $J^{\prime}_c \sim 0.135$. The different theoretical curves in Fig.~\ref{moir} correspond to 
different values of the ratio $r=J^{\prime}/J^{\prime}_c$.

The structure at low voltage $V$ of the differential conductance $G(V)=dI/dV$ is determined by the localized and conduction 
electrons of symmetry $\tau$ included in the model. 
We assume that the STM tip senses mainly the localized $3d$ states with some admixture of conduction states. 
Thus, the contribution of the model to $G(V)$ at zero temperature is given by Eq.~\ref{gm}, where the Green functions 
$G_{\tau \sigma }^{d}(\omega )$ depend only on $r$, and $q$ is a measure of the contribution of the conduction states.
In the experiment, there is also a linear background due to the contribution of other states, and $G_m(V)$ is affected by a 
factor $f$ which depends on the distance of the STM tip to the system. 
Therefore, to fit the experiment, the following expression is used
\begin{equation}
G(V)=f G_m(V)+A+BV,  
\label{g}
\end{equation}
which contains five parameters ($r, f, q, A$ and $B$), but the {\em shape} of each curve depends essentially on $r$, while 
$q$ controls the asymmetry. 
We have not included in the comparison with experiments the curve (a), which is similar to a rectangular dip formed by two 
step-like functions, because these steps are overbroadened in our calculations due to the limited resolution of the 
NRG at large energies~\cite{bulla08}.

The agreement between theory and experiment for cases (e) and (f) could be considerably improved by introducing intermediate
valence effects. This is in fact expected since a larger conduction density of states increases the parameter 
$\Delta= \pi v^2 \rho_c$ discussed in Section~\ref{nc}, which controls the degree of intermediate valence. 
It is well known that a smaller occupancy of the localized states shifts the Kondo peak to higher energies. 
Concerning case (e), the dip in the theory is more pronounced than in the experiment. 
However, as shown in Section~\ref{nc}, intermediate valence and finite temperature reduces the magnitude of the dip compared 
to that predicted by the A2CS1KM.

The comparison with the experiment can also be affected by the assumption of a constant density of conduction states and 
the effect of other orbitals not considered in our model. 
In any case, the experimental and theoretical results, including the LDA and NRG ones, strongly suggests that the underlying 
physics is that of the 2CS1AMA.

\subsection{Fe porphyrin molecules on Au(111)}

\label{porp}

Experiments similar to those for the Nc molecule on Cu(100) (see Section \ref{nc}), in which the STM tip is approached to the molecule, 
have been carried out for iron porphyrin molecules on Au(111)~\cite{meng23}. 
The LDA calculations indicate that the Fe spin is 1, with a partial occupancy of two orbitals (of symmetry $3z^2-r^2$ and $x^2-y^2$), 
that belong to different irreducible representations. 
In other words, the two channels are not equivalent. One expects that the A2CS1KM or the corresponding
Anderson model with non-equivalent channels, 
should be appropriate for the system.
In these experiments, the dip narrows as the contact regime is approached, but never turns to a peak, in contrast to Nc on Cu(100). 
Nevertheless, the results are consistent with our model assuming a weaker hybridization of the localized states with the substrate 
in comparison to Nc/Cu(100). 

\begin{figure}[h]
\begin{center}
\includegraphics[width=0.7\textwidth]{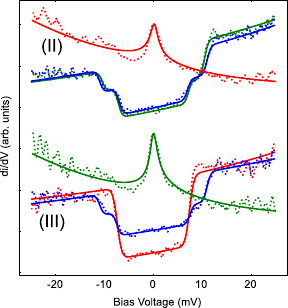}
\caption{(Color online) Differential conductance obtained at different positions (II and III) 
of a chain of three Fe-porphyrin-based molecules on top of herringbone-reconstructed Au(111) containing a Br atom. 
The presence of a peak (dip) signals a molecule in a ``Kondo'' (``spin-flip'') position.
The solid curves correspond to phenomenological fits of the data. 
Taken from Fig. 2 of \cite{gao23}. Copyright 2023 American Chemical Society.}.
\label{br}
\end{center}
\end{figure}

More recently similar experiments have been carried out using a Br decorated Au(111) surface~\cite{gao23}. 
In this case, depending on the particular position of the molecule with respect to the defects, 
sometimes a broad dip is observed and sometimes a narrow peak (see Fig.~\ref{br}), 
consistent with the A2CS1KM for parameters near the TQPT on the non-trivial and trivial topological sectors, 
respectively.

\section{Summary and discussion}

\label{summary} 

Since the seminal work of Nozières and Blandin~\cite{nozieres80}, the Kondo effect in situations involving 
multiorbital magnetic impurities and/or coupled to more than one conduction channels has become a highly active 
area of research within condensed matter physics. 
The primary motivation stems from the prediction of exotic electronic states, such as non-Fermi liquids 
(overscreened Kondo effect) and singular Fermi liquids (underscreened Kondo effect), exhibiting distinctive 
dynamic and thermal behaviors, that, on one hand, could shed light on unconventional physics in heavy fermion 
compounds, and, on the other hand, they serve as paradigmatic models for quantum many-body phenomena. 

The simplest generalization of the conventional Kondo effect to multiorbital systems, where a spin $S > 1/2$ 
is fully screened by $n=2S$ channels, has been relatively underexplored, perhaps because it was believed that no 
new physics could be found in this context. However, we have shown in a series of recent publications that 
the presence of single-ion magnetic anisotropy drives these systems through a topological quantum phase transition, 
that separates two topologically distinctive local Fermi liquids. 

Besides the theoretical relevance of our finding, we have shown that several systems consisting of isolated magnetic 
atoms or molecules on noble metal surfaces can be described by the 2-channel spin-1 Anderson model with anisotropy or 
its integer valent limit, the anisotropic two-channel spin-1  Kondo model. 
Both models exhibits the topological quantum phase transition, as a consequence of which the spectral density 
at zero temperature of the localized electrons has a jump between high values in the topologically trivial Fermi liquid 
phase to very low values in the non-trivial "non-Landau" Fermi liquid phase. Near the transition, the spectral density
is characterized by a narrow peak or dip (depending on the phase) mounted on a broad peak. 

It is worth clarifying that the topological transition studied in this work has a local character, as it pertains 
solely to the impurity's degrees of freedom. Furthermore, the topological properties arise from the frequency 
dependence of the impurity's Green's function, rather than from the momentum dependence, as is typically 
the case in the far more studied topological systems in condensed matter physics.

The abrupt transition might have technological applications. For example the system can act as a transistor, which is one of the most important components of an integrated circuit, as it acts as a switch by changing some parameter \cite{mathew18}. 
It also enlarges the field of single-molecule switches \cite{xu24} and bistable molecular systems \cite{xiao24}.

Several experiments with scanning-tunneling spectroscopy in different systems have in fact identified similar 
structures in the differential conductance. Five of these systems are listed in Section~\ref{comparison}. 
They were usually interpreted using different, more conventional theories, probably due to the fact of the novelty of 
the concepts related to the topological quantum phase transition. We expect that this work contributes to disseminate 
these rather novel ideas to the  community of condensed-matter and nanoscience researchers.

We have also shown that the concepts can be extended to larger spin and two-impurity systems.

\section*{Acknowledgments}

We thank Rok \v{Z}itko and Pablo Roura-Bas for several collaborations in subjects of this review.  
AAA acknowledges financial support provided by PICT 2018-01546 and PICT 2020A-03661 of the Agencia I+D+i, Argentina.
LOM and GGB acknowledge financial support provided by PIP 3220 of CONICET, Argentina.

\section*{ORCID iDs}

A A Aligia https://orcid.org/0000-0001-9647-3926\\
L O Manuel https://orcid.org/0000-0001-7793-1709\\
G G Blesio https://orcid.org/0000-0002-1962-0969

\end{document}